\documentclass[10pt]{article}
\renewcommand{\raggedbottom}

\usepackage{cite}

   \usepackage[dvips]{graphicx}
\usepackage[cmex10]{amsmath}

\usepackage{authblk}

\newcommand{\Frac}[2]{\displaystyle{\frac{\displaystyle{#1}}{\displaystyle{#2}}}}

\begin{document}
%
\title{Alya: Towards Exascale for Engineering Simulation Codes}

\author[1,2]{Mariano V\'azquez}
\author[1]{Guillaume Houzeaux}
\author[3]{Seid Koric}
\author[1]{Antoni Artigues}
\author[1]{Jazmin Aguado-Sierra}
\author[1]{Ruth  Ar\'{\i}s}
\author[1]{Daniel Mira}
\author[1]{Hadrien Calmet}
\author[1]{Fernando Cucchietti}
\author[1]{Herbert Owen}
\author[3]{Ahmed Taha}
\author[1]{Jos\'e Mar\'{\i}a Cela}
\affil[1]{Barcelona Supercomputing Center BSC-CNS, Spain}
\affil[2]{IIIA-CSIC, Spain}
\affil[3]{National Center for Supercomputing Applications-NCSA\\ University of Illinois at Urbana-Champaign, USA}
\affil[ ]{Contact: mariano.vazquez@bsc.es \\  guillaume.houzeaux@bsc.es \\ koric@illinois.edu}
\affil[ ]{\it Paper submitted to International Supercomputing Conference 2014}



\maketitle
\begin{abstract}
Alya is the BSC in-house HPC-based multi-physics simulation code. It is designed from scratch to run 
efficiently in parallel supercomputers, solving coupled problems. The target domain is engineering, 
with all its particular features: complex geometries and unstructured meshes, coupled multi-physics 
with exotic coupling schemes and Physical models, ill-posed problems, flexibility needs for rapidly including 
new models, etc. Since its conception in 2004, Alya has shown scaling behaviour in an increasing number of 
cores. In this paper, we present its performance up to 100.000 cores in Blue Waters, the NCSA supercomputer. 
The selected tests are representative of the engineering world, all the problematic features included: 
incompressible flow in a human respiratory system, low Mach combustion problem in a kiln furnace and coupled 
electro-mechanical problem in a heart. We show scalability plots for all cases, discussing all the aspects of such 
kind of simulations, including solvers convergence.
\end{abstract}

\maketitle

\section{Introduction}
Alya (see for instance \cite{alyaweb,GHouzeaux_MVazquez08a,Houzeaux2011e,eguzkitza2012parallel,lafo12}) 
is a simulation code developed at Barcelona Supercomputing Center (BSC-CNS) since 2004,
whose main architects are authors GH and MV. Alya is not a single-physics born sequential code, parallelized afterwards.
On the other hand, it was designed from scratch as a multi-physics parallel code. It's main features
are the following:
\begin{itemize}
\item It solves discretized partial differential equations (PDEs), preferring variational methods (particularly
Finite Elements).
\item Space discretization is based on unstructured meshes, with several types of elements (hexaedra, tetraedra,
prisms, pyramids... linear, quadratic...) implemented.
\item Both explicit and implicit time advance schemes are programmed.
\item Depending on the case, staggered or monolithic schemes are programmed. However, staggered schemes with coupling
iterations are preferred for large multi-physics problems.  
\item Parallelization is based on mesh partitioning (for instance using Metis \cite{metis}) and MPI tasks, which
is specially well-suited for distributed memory machines. On top of that, some heavy weight loops are parallelized 
using OpenMP threads. Both layers can be used at the same time in a hybrid scheme.
\item Alya sparse linear algebra solvers are specifically developed, with a tight integration with the overall 
parallelization scheme. There are no third-parties solver libraries required. 
\item Alya includes some geometrical tools which operate on the meshes for smoothing, domain decomposition or mesh sub-division.
In particularly, the latter is a key tool for large-scale simulations \cite{HouzeauxMultipli}.
\end{itemize}

This paper addresses the performance of Alya in supercomputers, running up to 100K cores in Blue Waters,
the sustained peta-scale system \cite{bluewatersweb} hosted at the University of Illinois'
National Center for Supercomputing Applications (NCSA). Blue Waters consists of traditional Cray XE6 compute nodes 
and accelerated XK7 compute nodes in a single Gemini interconnection fabric. Only XE6 nodes where used in this work, 
with each node containing two AMD Interlagos processors which totals to 16 floating point cores/XE6 node (NCSA, USA).
Performance is measured through scalability when simulating coupled multi-physics problems in complex geometries coming
from different domains. The selected cases are incompressible flow in the respiratory system, turbulent 
low Mach incompressible flows with combustion in a kiln furnace and non-linear solid mechanics coupled with electro-physiology
in a human heart. 

\section{Alya general view}
Alya is organized in a modular way: {\it kernel}, {\it services} and {\it modules}, which can be separatedly compiled and linked.
Each {\it module} represents a different set of Partial Differential Equations (PDE), i.e. each module is a {\it physics}.
To solve a coupled multi-physics problem, all the required modules are active and interacting following a certain workflow.
Alya's {\it kernel} controls the run (it contains the solvers), the input-output and everything related to the mesh and geometry.
With kernel and modules, a given Physical problem can be completely solved. The {\it services} are supplementary stuff,
notably the parallelization service. Kernel, modules and services have well-defined interfaces and connection points. 

\subsection{Computational Mechanics Equations: the theoretical setup}

Generally speaking, Alya deals with Computational Mechanics problems that can be modelled through
conservation laws expressed as a set of partial differential equations:

\begin{eqnarray*}
\partial^*_t{\Phi^{\alpha}} = \partial_{x_i}{F^{\alpha}_{i}} = \partial_{x_i}{C^{\alpha}_{i}} + \partial_{x_i}{K^{\alpha}_{i}}
\end{eqnarray*}

where $\Phi^{\alpha}$  is the $\alpha$-equation of the set. $F^{\alpha}_{i}$ is the compact notation of the fluxes
for each of the equations, being divided in two terms, $C^{\alpha}_{i}$ and $K^{\alpha}_{i}$, for convenience.
The temporal derivative $\partial^*_t{\Phi^{\alpha}}$ is starred to note that it can be of first (like in fluid flows or
excitable media) or second order (like in solid mechanics or acoustics). 
Subindices run through the space dimension of the space 
domain $\Omega$. To these equations, boundary and initial conditions must be added 
depending on the problem under study.

The variational form is obtained by projection on a space $W$ with its usual
properties, where $\forall \; \Psi \in W$ after integration by parts of
some of the fluxes (those labelled with $K$) yields, we obtain

\begin{eqnarray}
\partial^*_t{}\int\Psi^{\alpha}\Phi^{\alpha}d\Omega = \int\Psi^{\alpha}\partial_{x_i}{C^{\alpha}_{i}} d\Omega
- \int\partial_{x_i}{\Psi^{\alpha}}K^{\alpha}_{i} d\Omega \nonumber \\
+ \int\Psi^{\alpha}K^{\alpha}_{i} n_i d\partial\Omega.
\label{eq:weakcompact}
\end{eqnarray}

The last of the right hand side terms can be used for imposing Neumann-like boundary conditions on the
fluxes themselves, being $\partial\Omega$ the domain boundary and $n_i$ its exterior normal vector.
The interpolation space where the variational form solution is to be found is (practically) the
same as $W$, explicitly time independent. For that reason, the time derivative has been taken out of the space
integral, with the additional hypothesis that $\Omega$ is not changing with time.
These equations govern problems in fluid mechanics, solid mechanics, chemical reactions, quantum
mechanics, heat transfer, etc.

In Alya, the method used to discretize in space the weak form Equation \ref{eq:weakcompact}
is the finite element method.  For the cases that additional numerical
stabilization is required, we usually follow the the so-called Variational MultiScale
method (see for instance \cite{GHouzeaux_JPrincipe08}).  Time is discretized
using finite differences to obtain either explicit or implicit schemes of different orders.  Time
and space discretization of the weak form leads to a $\mathbf{A} \mathbf{u} = \mathbf{b}$
algebraic system.

The equations terms contain all the ``physics'' of
a given problem. The governing equations are discretized in time and space in a certain way and programmed
in a module. So the module main task is to compute the elementary matrix and righ-hand-side of its corresponding
set of equations, including all the numerical subtleties, boundary conditions, material models and so on.
These matrices are assembled in a global matrix and a global right-hand-side vector, creating an algebraic
system.
 
\begin{figure}[!t]
  \centering
  \includegraphics[width=3.5in]{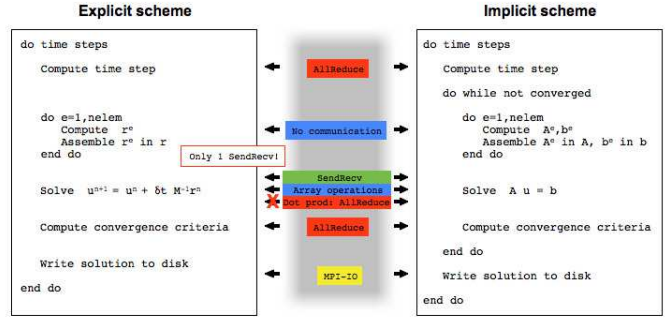}
  \caption{Typical explicit and implicit schemes}
  \label{fig:exp-imp}
\end{figure}

These algebraic systems can be very large. In Alya, we prefer to solve them using iterative methods
which are very well suited for parallel programming. According to the problem solved, we follow
a wide range of iterative strategies. On one hand, the explicit schemes, which can be viewed as
the most simple iterative method with a unique simple (Richardson) iteration per time step.
On the other hand lie schemes such as GMRES, BiCGSTAB, CG \cite{YSaad03} or Deflated 
CG \cite{YSaad00,RLohner_FMut10}. Both explicit and implicit schemes are illustrated in 
Figure \ref{fig:exp-imp}. Apart from the time loop,
a linearization loop may be necessary for non-linear problems when using the implicit scheme.

\subsubsection{Coupling schemes}

In Alya, Multi-physics problems are simulated through coupling modules. Although the code structure allows much more
complex situations, in this paper we focus in a certain
kind of coupling, where all happens on the same domain and domain discretization. In these cases, the main features are:
\begin{itemize}
\item All problems are solved on the same mesh, no interpolation is required.
\item The total number of degrees of freedom is equal to the number-of-nodes {\it times} the total number-of-variables that
defines de problem. Each module covers a set of variables. For instance, when incompressible flow is coupled with thermal
transport, pressure, velocity and temperature define the problem. 
\item However, Alya never has the large corresponding matrix system, because the multi-physics solution scheme 
is a staggered one. For each timestep, the block corresponding to each module is solved sequentially. If required,
coupling iterations can be performed to increase accuracy and/or robustness. When more than two modules are coupled,
the iterative scheme can be very complex.
\item Scalability is measured based on the total time for all the modules as they run in an increasing number of processors. 
If a module is not properly scaling it will spoil the total performance.
\end{itemize}

\subsection{The parallelization layer}
 The parallelization paradigm in Alya is a sub-structuring method, using a Master-Worker interaction model between the CPUs. 
 Sub-structuring methods consist essentially in distributing the work among the Workers, letting the Master in charge of 
 simple tasks like I/O. In that sense, most of the iterative solvers implemented in Alya are parallelized classical
 solvers like GMRES or Conjugate gradient, which convergence do not depend on the number of CPUs. Preconditioners
 using coarse space corrections (Multigrid, Deflated Conjugate gradient) are implemented independently of the 
 number of CPUs as well. This is not a restriction but a deliberate decision. 
 In fact, if one has tuned a set of parameters to achieve convergence on a given number of CPUs, one would like to obtain 
 the same convergence using more CPUs. This is a crucial point when considering industrial simulations where the user usually 
 does not necessarily have time to try different sets of parameters depending on the number of available CPUs.
 Nevertheless, domain decomposition methods like Additive Restricted Schwarz (RAS), Block LU (one block per subdomain),
 Schur complement solvers, as well as subdomain dependent preconditioners like linelet \cite{linelet2} are also implemented in Alya. 
 These solvers and preconditioenrs are generally used whenever classical solvers appear not to be robust enough.
 
 \subsubsection{Master-Worker strategy}
 
 All the details on the parallelization of Alya can be found in 
 papers such as \cite{GHouzeaux_MVazquez08a,Houzeaux2011,RLohner_FMut10},
 we give here the general idea. The parallelization is based on a Master-Worker strategy.
 The Master reads the mesh and performs the partition of the element graph with METIS \cite{metis}, an 
 automatic mesh partitioner which balances the number of elements while minimizing the subdomain
 boundary/interface surfaces, that is the communications. For the pure MPI strategy, each core will be in charge of 
 each subdomain, which are the workers. The workers build the local matrices ($\mathbf{A}_i$) and right-hand side 
  ($\mathbf{b}_i$), and are in charge of the resulting system solution in parallel. In the assembling tasks, very
  few communications are needed between the workers and the scalability only depends essentially on the load balancing. 
  Basically only few ${\tt MPI\_AllReduce}$ are required to compute solution residual, critical time step, etc. 
  In the iterative solvers, the scalability depends not only the load balancing but also on the size of the boundaries 
  between the subdomains and on the communication scheduling.

\subsubsection{Communication types and scheduling}

In the iterative solvers, two main types of communications are usually needed. 
\begin{itemize}  
   \item Global communications via ${\tt MPI\_AllReduce}$, which are used to compute residual norms,
         time steps and scalar products involved in algebraic solvers;
   \item Point-to-point communications via ${\tt MPI\_ISend}$ and ${\tt MPI\_IRecv}$, which are used in 
         algebraic solvers when sparse matrix-vector (SMV) products are needed. 
\end{itemize}

 We mentioned earlier that the parallelization of Alya is based on a sub-structuring approach in which most of
 the solvers and preconditioners are implemented independently of the number of subdomains.
 Therefore the parallel solution is, up to round off errors, the same as the sequential one at any moment because mesh
 partition is only used for distributing work without changing the sequential algorithm. 
 Figure \ref{fig:exp-imp} shows these two types of communications in explicit and implicit schemes. 
  The element loop consists of the local (to each subdomain) matrix and RHS assemblies and does not involve 
  communication. Therefore, the parallel performance of the explicit scheme is expected to be dominated by the
  load balance.

\begin{figure}[!t]
  \centering
  \includegraphics[width=3.3in]{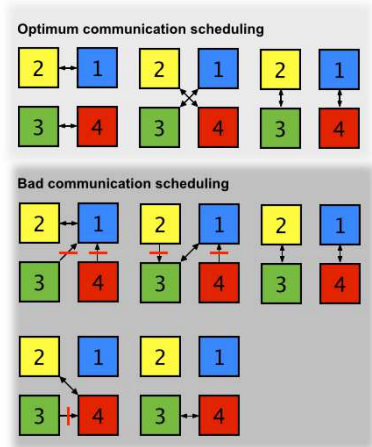}
  \caption{Left: Scheduling strategy on a simple example.  Top, optimum
           and communication scheduling in 3 steps. Bottom, Bad communication scheduling
           in 5 steps.}
  \label{fig:schedul}
\end{figure}

 Another key issue of communication is scheduling \cite{PBrucker03}. Figure
 \ref{fig:schedul} shows the kind of problem that can arise when data
 transfer is not properly scheduled. In this case, four subdomains have to exchange data
 with all the others. The optimum scheduling is shown on the top part of the figure. 
 On the bottom part, no scheduling is used and subdomains try to exchange their
 boundary data in a lexical order. In the first communcation step, subdomains
 3 and 4 cannot send their data to subdomain 1, as this one is being exchanging
 data with subdomain 2. A bad scheduling can strongly penalize scalability.

\subsubsection{Data structure}
\begin{figure}[!t]
  \centering
  \includegraphics[width=3.5in]{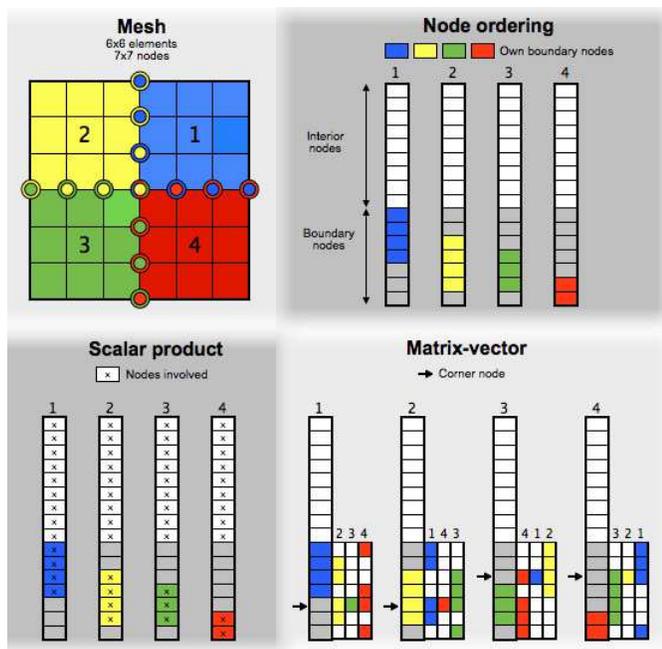}
  \caption{Organization of data in Alya and types of communications.}
  \label{fig:smv}
\end{figure}

A specific data structure for distributed memory parallelization has been used
and can be briefly explained through a simple example illustrated in Figure
\ref{fig:smv}.  It shows an example of mesh partitioning into
four subdomains (top left) and its corresponding node numbering (top right). In
each slave, interior nodes are first ordered. Then, boundary nodes (grey) are
divided into {\em own boundary nodes} and {\em others bounday nodes}.  
The tag {\em own boundary nodes} cannot be
repeated in more than one subdomain and are obtained by partitioning the
 subdomain boundary with METIS. The {\em own boundary node} definition is useful when scalar
 products are needed to avoid repeating the contribution of boundary nodal values. The nodes
involved in scalar products are shown in Figure \ref{fig:smv} (right)
with a $\times$ sign, being 13 nodes in this particular case (bottom left).
Finally, the nodes involved in the ${\tt MPI\_ISend}$ and ${\tt MPI\_IRecv}$
after a SMV are shown (bottom right).

\subsubsection{Asynchronous and scheduled SMV}

Using the data structure and the scheduling introduced earlier, the Sparse Matrix Vector product
is computed together with a non-blocking send-receive as follows:
\begin{enumerate}
\item Perform local SVM product $\mathbf{y}_i = \mathbf{A}_i \mathbf{x}_i$ on boundary nodes;
  \item Exchange $\mathbf{y}_i$ with neighbors using non-blocking ${\tt MPI\_ISend}$ and ${\tt MPI\_IRecv}$
        according to the communication scheduling;
   \item Perform local SVM product $\mathbf{y}_i = \mathbf{A}_i \mathbf{x}_i$ on interior nodes;
  \item Synchronize the solution updates with ${\tt MPI\_WaitAll}$.
\end{enumerate}

\section{The examples}
In this paper we analyze Alya's parallel behaviour through three different examples: 
\begin{itemize}
\item The human respiratory system: transient incompressible flow.
\item The kiln furnace: transient incompressible flow with a low-Mach approach, heat transport and chemical
reactions.
\item The electro-mechanical cardiac model: transient non-linear solid mechanics with an hyperelastic model and
excitable media. 
\end{itemize}
In all cases we start with meshes in the range of a few million elements, which are progressively subdivided 
in parallel using the {\em mesh multiplication} algorithm described in \cite{HouzeauxMultipli}, in order 
to produce large enough meshes to feed a supercomputer up to a hundred thousand cores.
 Before analyzing the parallel performance, we proceed to briefly describe each of the examples and
 their associated numerical strategies.

\subsection{The respiratory system}

Computational simulation enables the mechanics of respiratory airflow to be explored in detail, 
with considerable potential benefits for healthcare and protection. Resolving the complex time-dependent 
flow in the large airways poses a severe challenge. Various compromises are generally made, such as 
restricting the portion of the airways considered and approximating the flow conditions or physics. In this
example, the unsteady flow in a subject-specific model of the domain that extends from the face to 
 the third branch of the bronchopulmonary tree is simulated.

The whole airway geometry was defined from a single subject, identified via retrospective examination of CT
images obtained from clinical records at St Marys Hospital, Paddington. Consent was obtained to use this 
data as the basis for airway segmentation and reconstruction. Segmentation of the airways was performed 
using the Amira package (TGS Europe) and required some manual intervention, particularly in the nasal 
airways. There the fine bone structure challenges the resolution typical of data acquired under routine 
clinical protocols, but the fidelity of the reconstructed data was carefully checked by ENT surgeons. 
Translation of the coarse segmentation into a smooth surface was performed using in-house, curvature 
adapted smoothing procedures. Mesh generation was accomplished in stages, 
 using the Gambit and TGrid packages (Ansys Ltd.). This work was done in collaboraiton with D. Doorly
 and A. Bates from Imperial College (UK).

The solution of this problem involves the solution of the incompressible Navier-Stokes
equations. The time discretization is based on a second order BFD scheme and the linearization is carried
out using the Picard method. The space discretization is based on the variational
multiscale method (VMS) and is extensively described in \cite{GHouzeaux_JPrincipe08}. 
At each time and linearization iteration, the following system 
\begin{eqnarray} 
   \left[ \begin{array}{ll}
          \mathbf{A}_{uu} & \mathbf{A}_{up} \\
          \mathbf{A}_{pu} & \mathbf{A}_{pp}
   \end{array} \right]
   \left[ \begin{array}{l}
          \mathbf{u} \\
          \mathbf{p}
   \end{array} \right]
   =
   \left[ \begin{array}{l}
          \mathbf{b}_u \\
          \mathbf{b}_p
   \end{array} \right]
\end{eqnarray}
is solved, where $\mathbf{u}$ and $\mathbf{p}$ are velocity and pressure nodal unknowns. The direct solution of
the system is usually referred to as monolithic scheme. In order to avoid the use of complex preconditioners 
to account for the velocity-pressure coupling involved in this monolithic system, an algebraic fractional scheme was
in developed \cite{Houzeaux2011}.
This scheme enables to segregate the solutions of the velocity and pressure at the algebraic level, 
by solving the pressure Schur complement using an iterative method (herein the Orthomin(1)). 
This strategy offers two main advantages. Firstly, with respect to the
monolithic scheme, one shot of the method involves the solution of the momentum equation and the solution of 
a symmetric system for the pressure (Laplacian) representing the continuity equation. The momentum equations usually 
converge very well, even with a simple diagonal preconditioner. The continuity equation is solved with the Deflated 
Conjugate Gardient solver (DCG) \cite{RLohner_FMut10}, together with a linelet preconditioner when anisotropic 
boundary layers \cite{linelet2} are present. Secondly, with respect to classical fractional step methods, no fractional errors
are introduced and the solution of converges to the same as the monolithic one.
 What is important to note here is that the solution strategy which consists
 in solving the pressure Schur complement instead of attacking directly the monolithic scheme, should be understood as part
 of the algebraic solution strategy of the Navier-Stokes system, and not a fancy trick to escape from this scheme.

\begin{figure}[!t]
  \centering
  \includegraphics[width=3.5in]{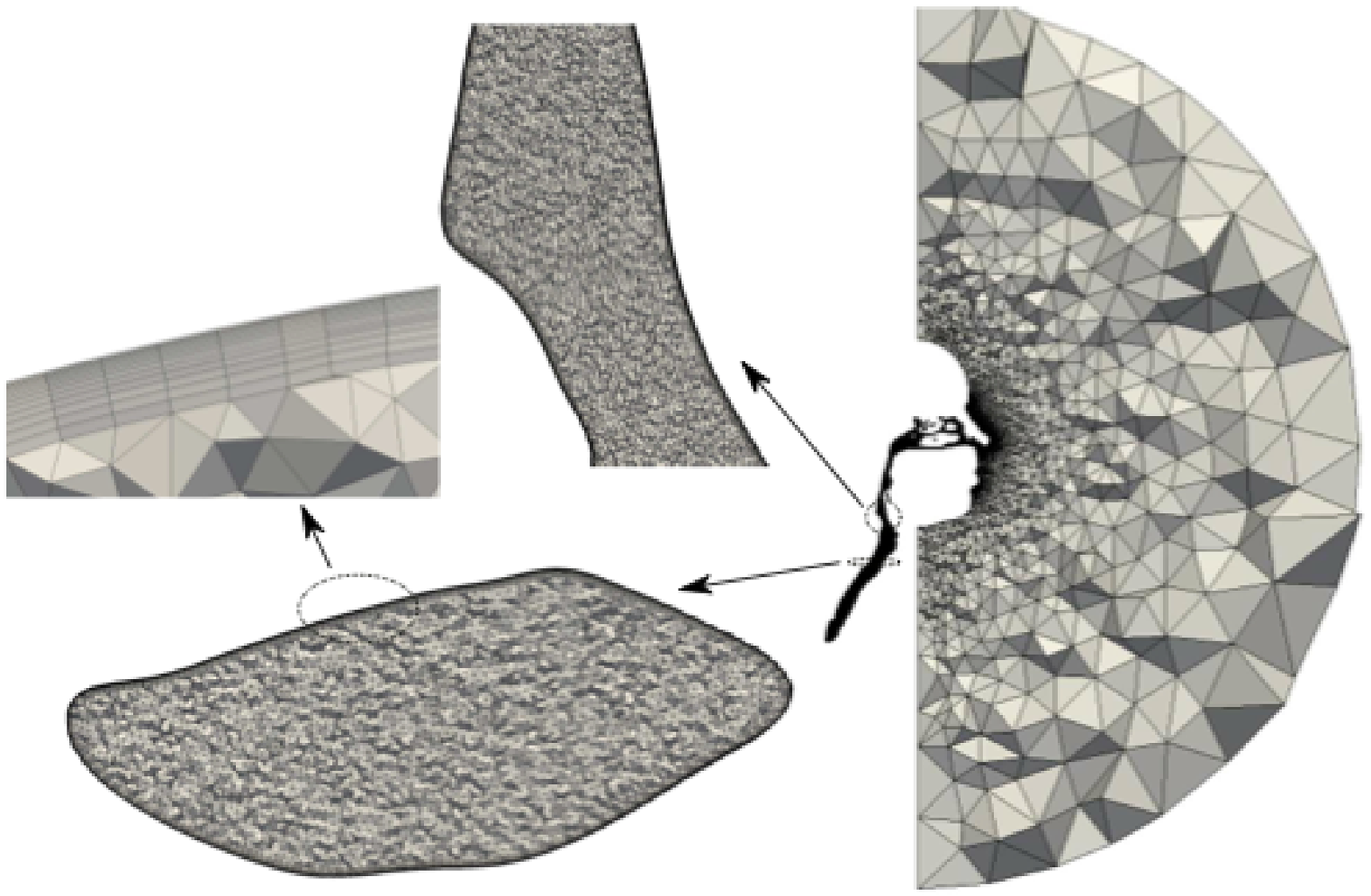}
  \includegraphics[width=3.5in]{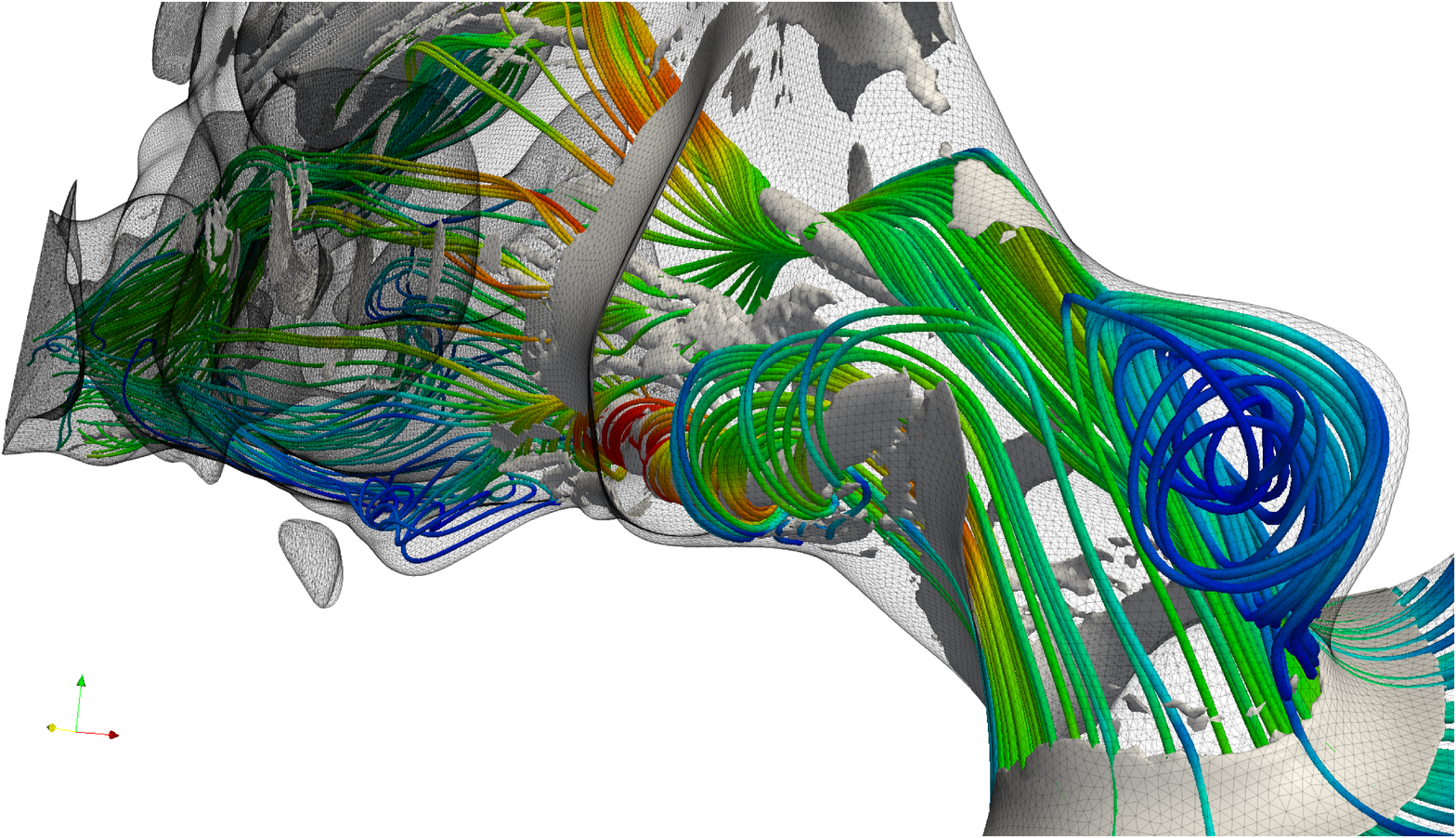}
  \caption{Respiratory system initial mesh (top) and close up view with streamlines.}
  \label{fig:respira} 
\end{figure}

\begin{figure}[!t]
  \centering
  \includegraphics[width=4.5in]{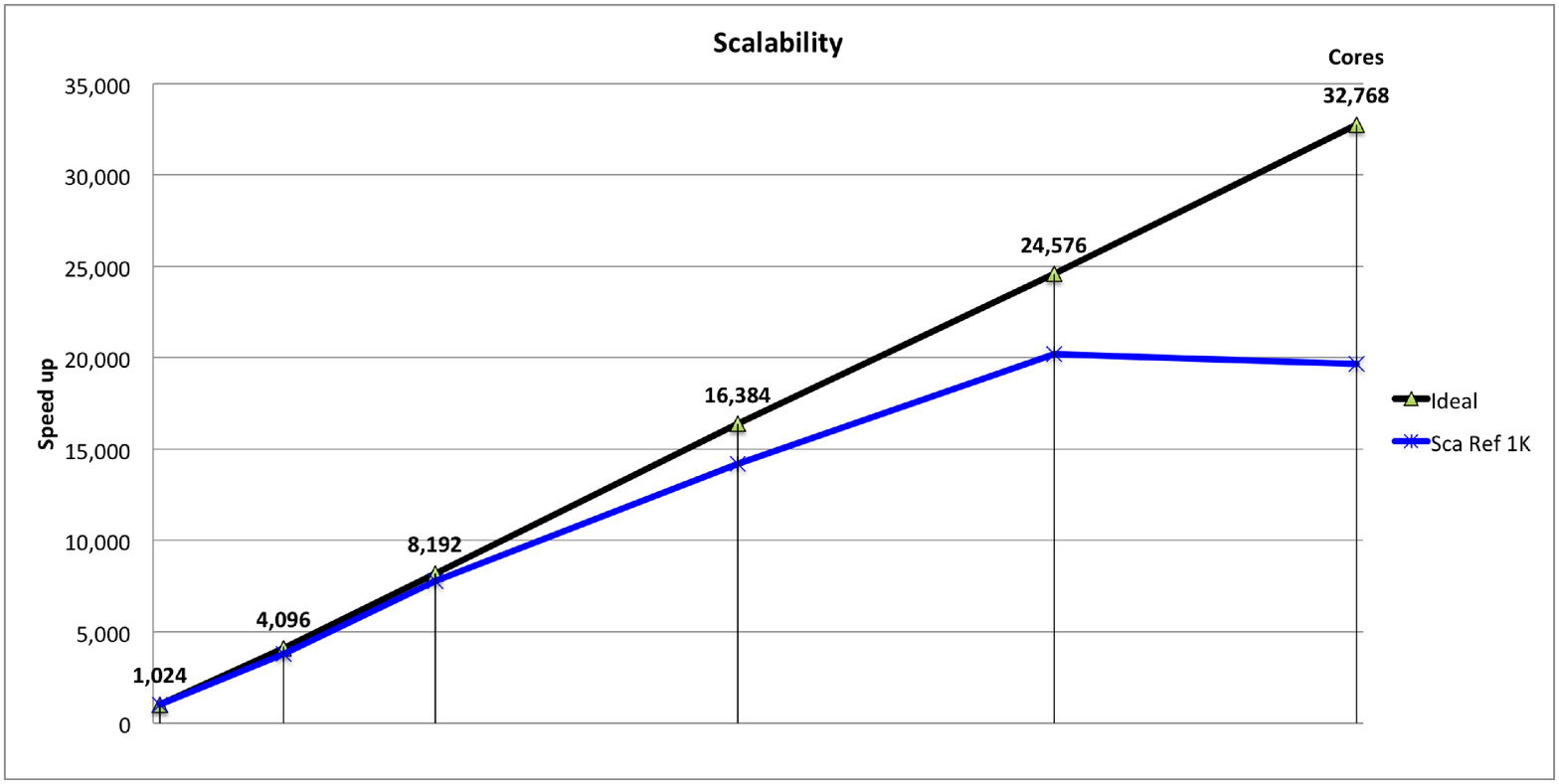}
  \vskip0.2cm
  \includegraphics[width=4.5in]{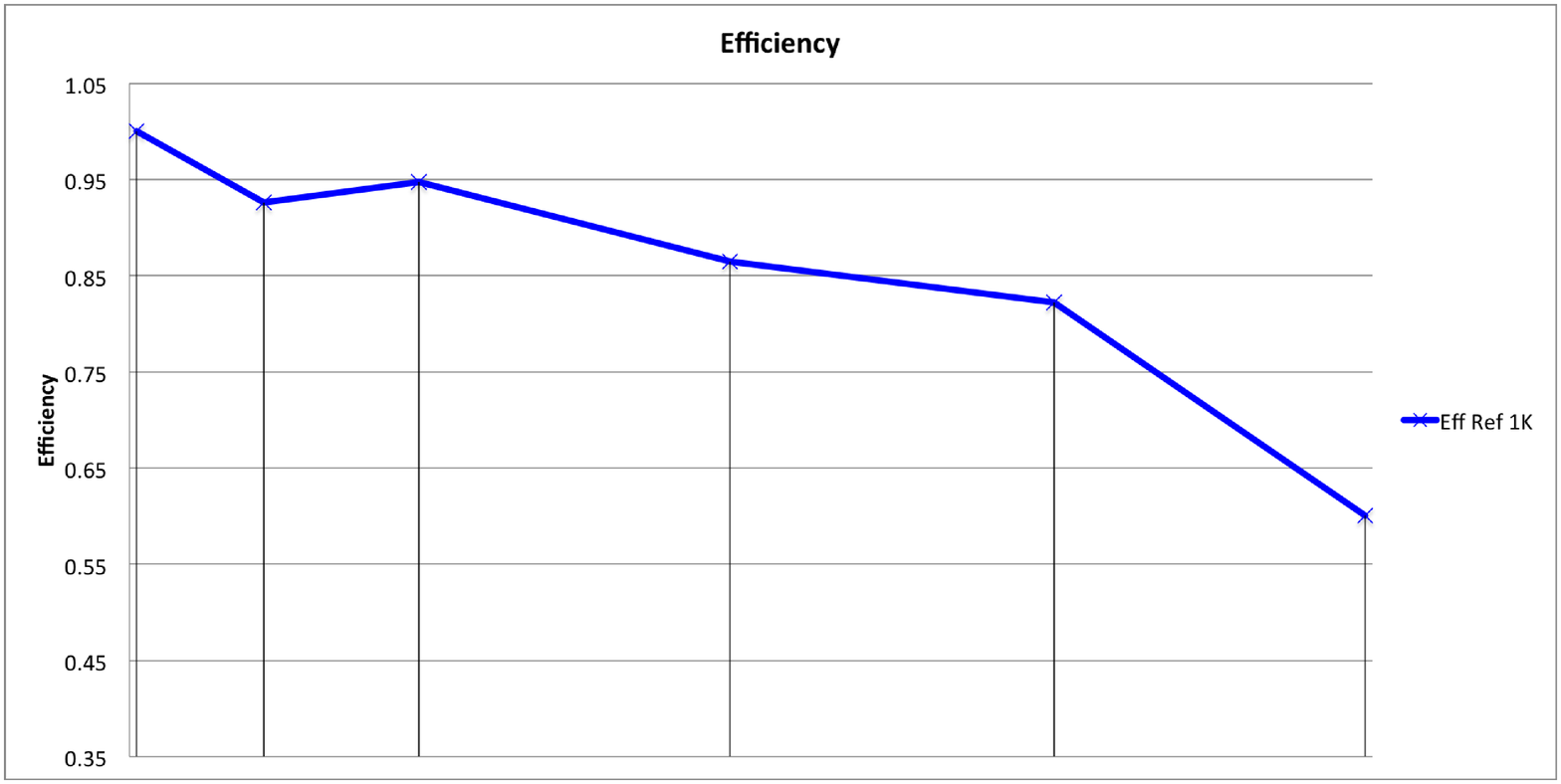}
  \vskip0.2cm
  \includegraphics[width=4.5in]{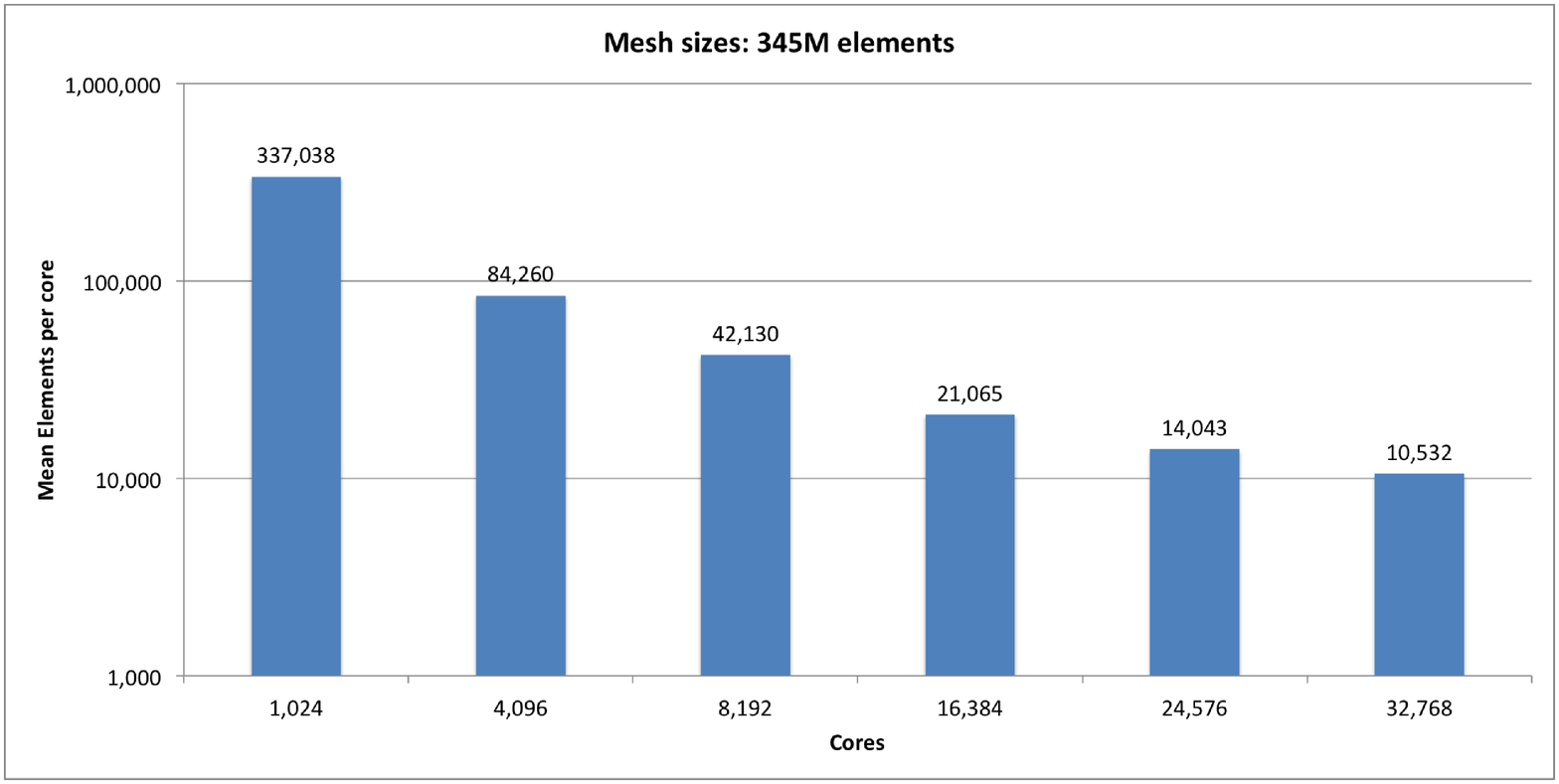}
  \caption{Scalability, efficiency and mean-elements-per-core for the respiratory system simulation.}
  \label{fig:respi-scalabi}
\end{figure}

Figure \ref{fig:respi-scalabi} shows the strong scalability and efficiency for the respiratory system problem.
Scalability is measured comparing the CPU time taken to solve one simulation time step in an increasing number
of processors. Efficiency higher than 0.80 is sustained up to 24K processors, where it starts to be degraded
due to the higher ratio of communications / computing time. The bars plot at the botton gives the mean-elements-per-core
figure. This is very useful to establish a sweet spot, which depending on the problem physics and size, sets the number
of processors you need to maintain a high efficiency. If we choose 0.80 as the limit, for this case the sweet spot
is around 15.000 elements per core. 

\subsection{The kiln furnace}
At the heart of the cement production process lies the kiln, a
tilted rotary oven where raw materials are heated to reaction temperatures
to form small pellets called clinker, which is then ground to make cement
powder \cite{Mujumdar2006}. In addition to reducing its raw material consumption, improving
the efficiency of cement kiln is a main concern of the industry, as the kiln is
the main consumer of energy in the production process.
A kiln is a tilted cylindrical vessel rotating at a fixed frequency of about 5
rpm. Its length ranges between 50 and 180 meters, and its diameter between
2 and 4 meters. On the high part of the kiln the raw material is fed in,
sometimes as a dry powder and others as a wet sludge, where it begins the
process of clinkerization. On the lower part of the kiln a large burner ejects
fuel, typically pulverized coal or waste material. The burner has a primary
air injector, and secondary and tertiary injectors 
that add swirl motion to the flow acting as a flame stabilization mechanism, 
which can be up to 10 meters long. The walls of the kiln are a mixture of refractory bricks and metal. 
Furthermore, cement stuck to the walls around the lower third of the oven forms a coating shell 
critical for operation of the kiln.

In this example, the gaseous phase of a rotary kiln is simulated using large-eddy simulation (LES). 
The numerical scheme to solve this coupled problem 
is based on a staggered algorithm that solves the Navier-Stokes equations at the low-Mach limit, 
the enthalpy transport equation expressed in terms of temperature along with the transport and reaction 
of the chemical species. In this case, we consider six chemical species to represent the oxidation of methane.

The flow equations are solved using 
a second order backward difference sheme (BDF) with
a Newton-Raphson linearization method. The momentum 
and continuity equations are solved with unsymmetric and symmetric
iterative solvers respectively. For the momentum equations, the GMRES
is considered while the Conjugate Gradient (CG) or Deflated CG are the choices for the
continuity equation \cite{RLohner_FMut10,GHouzeaux_MVazquez08a}. 
The GMRES solver is also employed to solve
for the enthalpy, turbulence quantities 
and species mass fractions. The Gauss-Seidel
iterative method is employed to 
solve the species mass fraction until the targeted
convergence.

\begin{figure}[!t]
  \centering
  \includegraphics[width=3.5in]{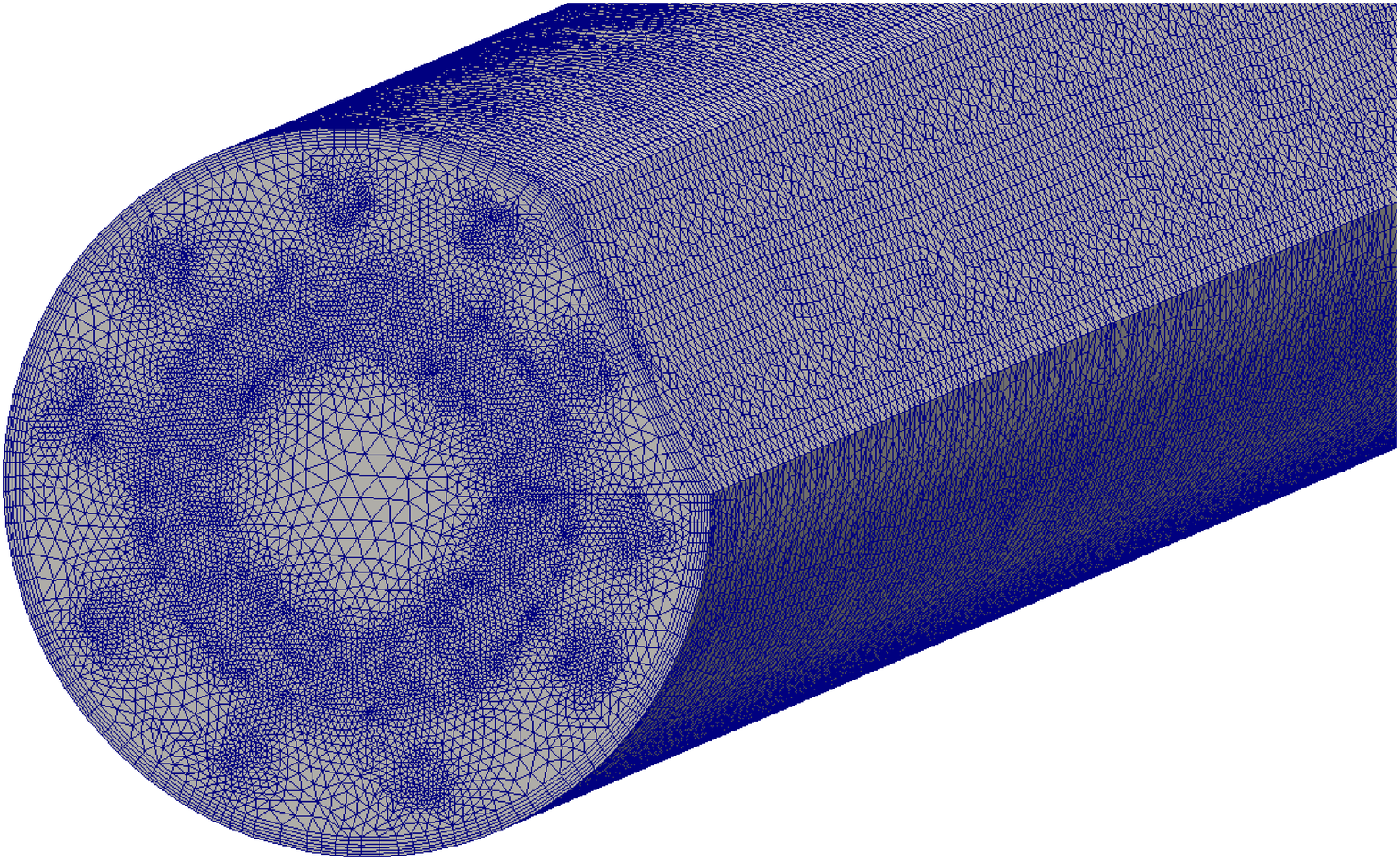}
  \includegraphics[width=3.5in]{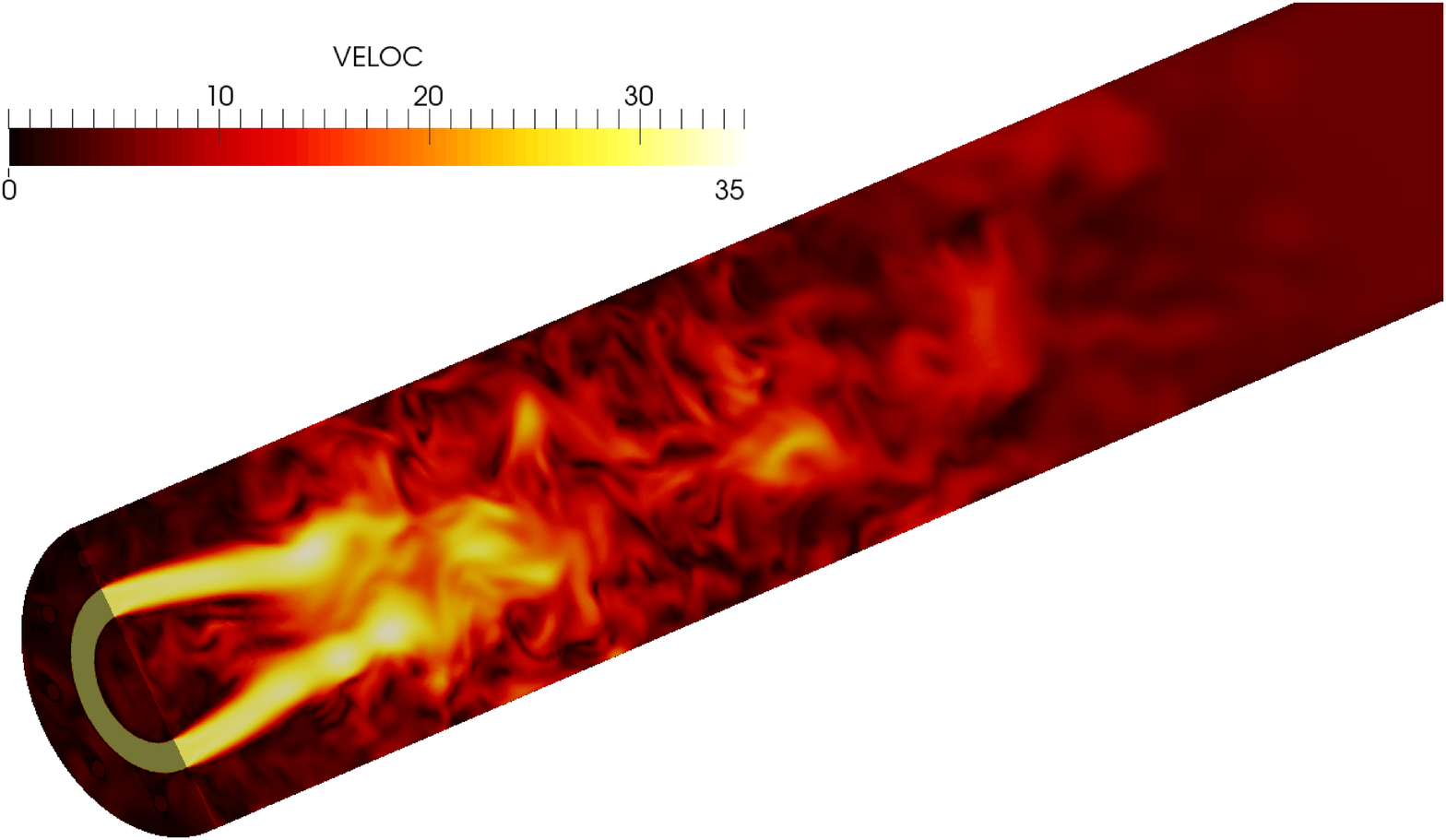}
  \caption{Kiln initial mesh close-up (top) and cut with velocity contours.}
  \label{fig:kiln} 
\end{figure}

\begin{figure}[!t]
  \centering
  \includegraphics[width=2.5in]{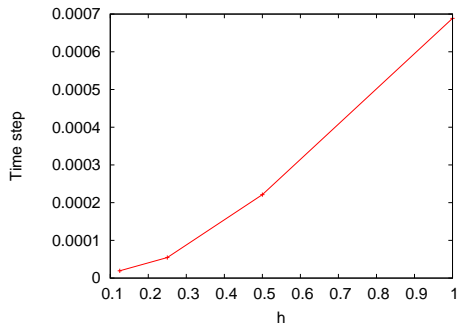}
  \caption{Kiln. Time step values for the four meshes.}
  \label{fig:time-step} 
\end{figure}

In this problem and following \cite{HouzeauxMultipli}, 
four different levels of mesh subdivision have been considered, 
referred to here as $h=1,\,1/2,\,1/4,\,1/8$, from the coarsest to the finest, respectively. The numbers
of elements are $8.25M$, $66.0M$, $528M$ and $4.22B$, respectively.
The time step size is computed as a mulitple of the critical time step. Figure \ref{fig:time-step} shows
the average time step values of the first ten time steps, computed for the four meshes.

\begin{figure}[!t]
  \centering
  \includegraphics[width=4.5in]{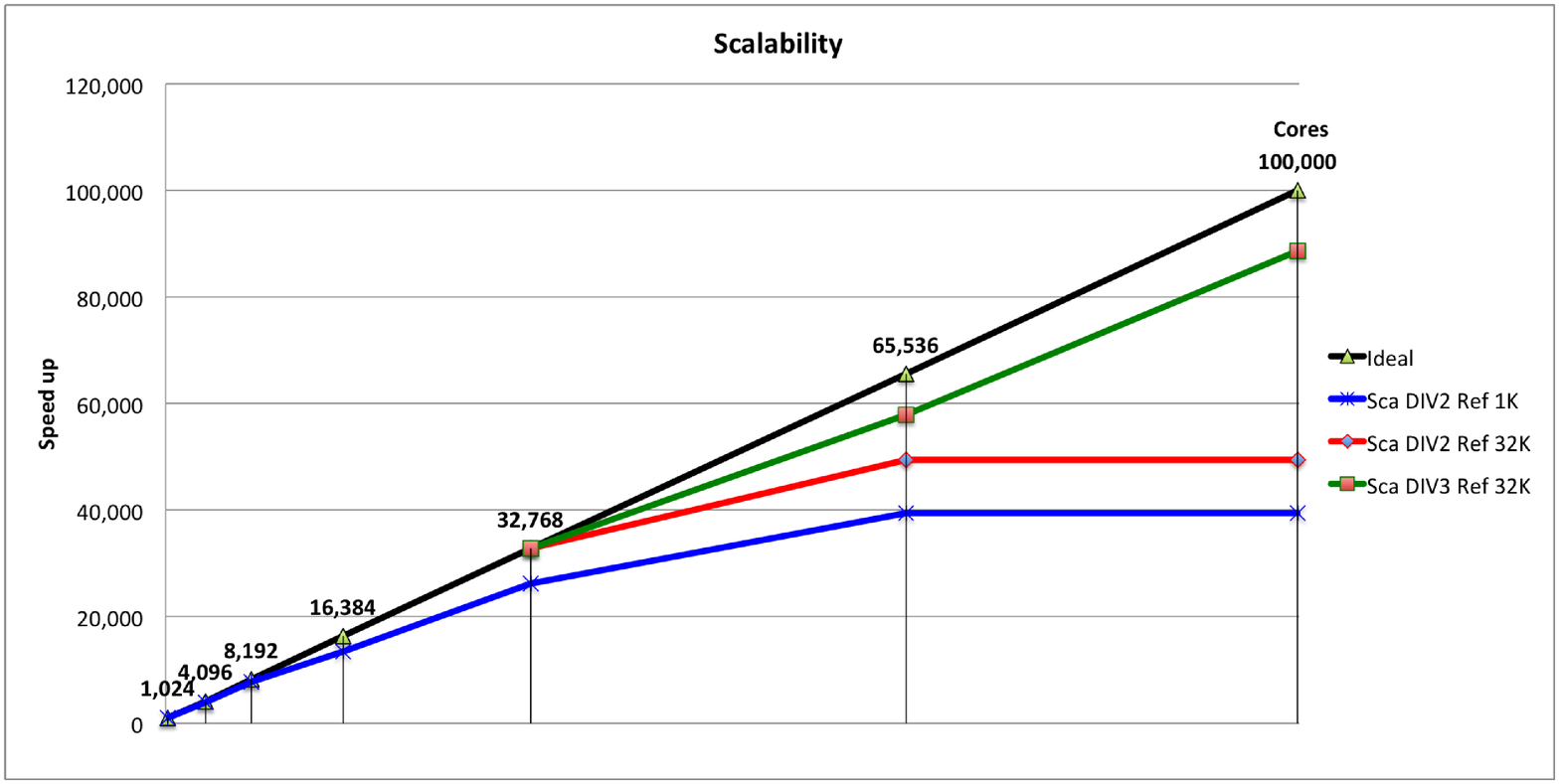}
  \vskip0.2cm
  \includegraphics[width=4.5in]{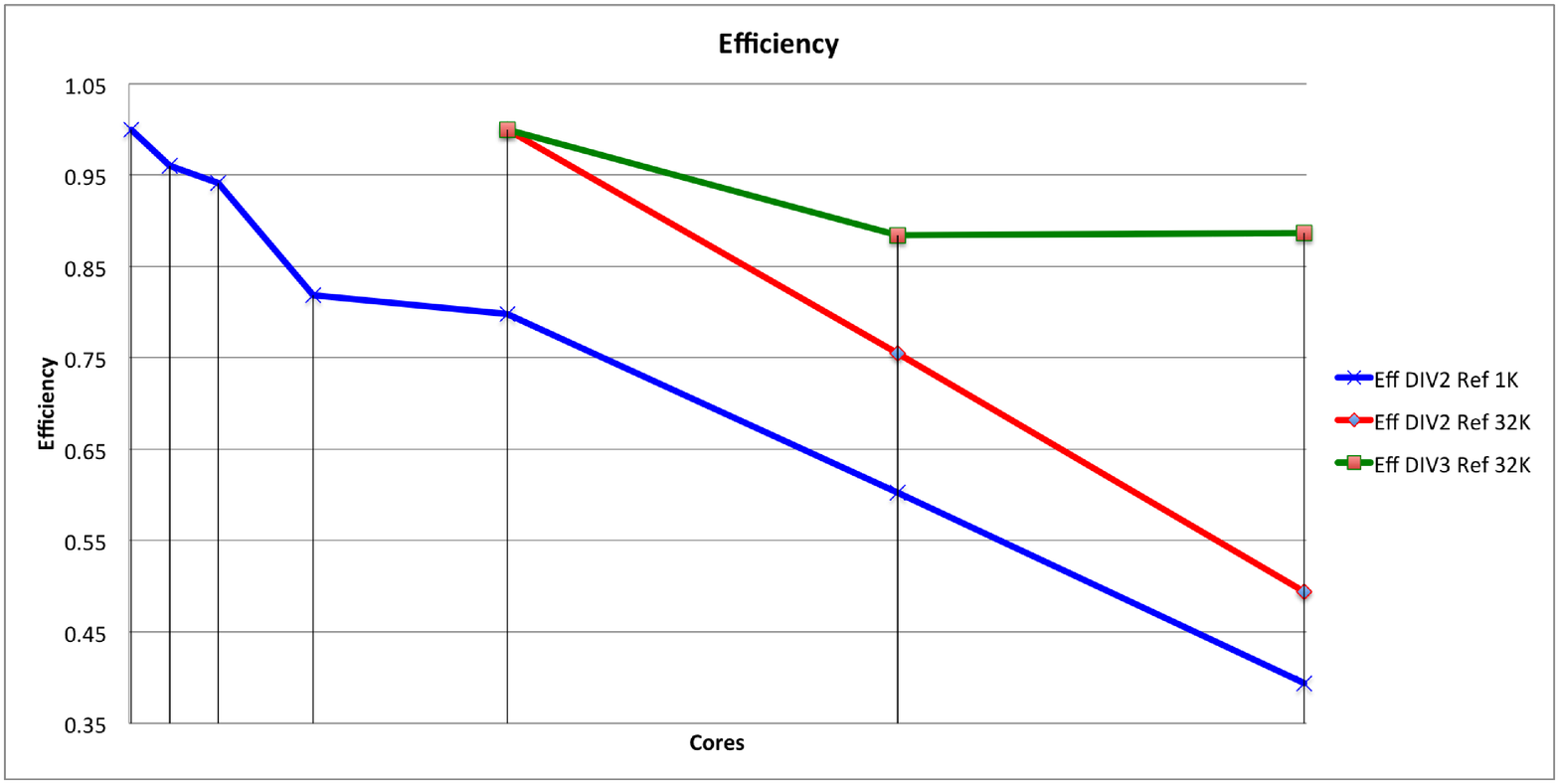}
  \vskip0.2cm
  \includegraphics[width=4.5in]{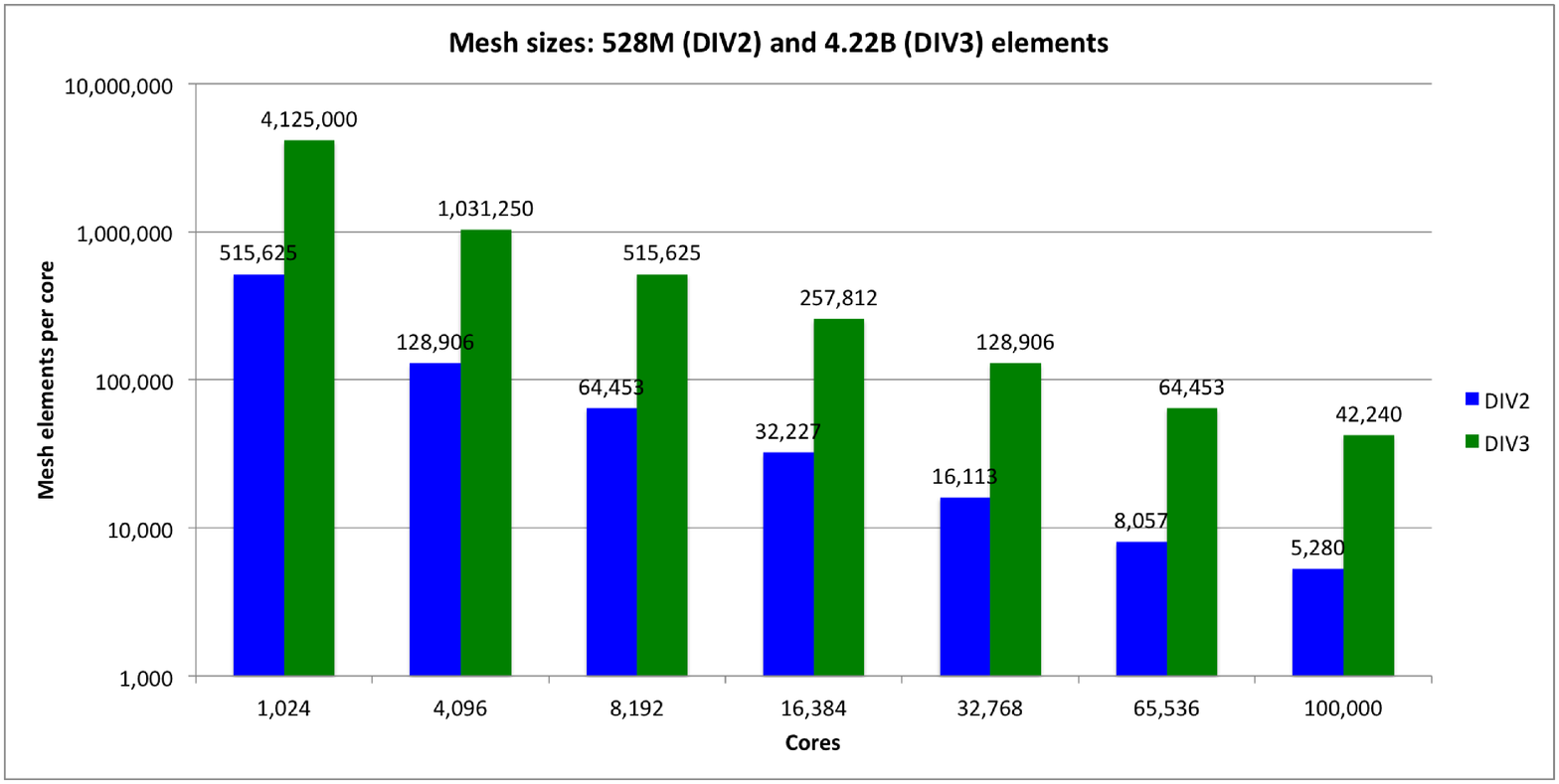}
  \caption{Scalability, efficiency and mean-elements-per-core for the kiln furnace simulation.}
  \label{fig:kiln-scalabi}
\end{figure}

Figure \ref{fig:kiln-scalabi} shows the strong scalability and efficiency for the kiln example. The plots show
the total scalability, measured summing up the CPU times for all the Physical problems solved, namely
low Mach, temperature and chemical reactions. In this example
we show the results for two meshes: 528M (called DIV2) and 4.22B elements (called DIV3). For DIV2 (labelled ``DIV2 Ref 1K'')
the scalability
is measured all the way from 1024 up to 100K cores, with the sweet spot around 16K mean elements per core. Beyond
that point, efficiency falls below 0.80. For DIV3, i.e. the largest mesh,  
we run the last three points of the plot, 32768, 65536 and 100000 cores,
using as the scalability normalizing value the CPU time obtained for 32768 (labelled ``DIV3 Ref 32K''). 
In order to be fair with the comparison,
we have added the scalability and efficiency plots for DIV2, but now normalizing with 32768 instead of 1024 (labelled ``DIV2 Ref 32K''). 
As expected, ``DIV2 Ref 32K'' is very close to a translation upwards of ``DIV2 Ref 1K''. On the other hand, ``DIV3 Ref 32K''
presents a much better scalability and efficiency, with a sustained large efficiency up to 100.000 cores.

\begin{figure}[!t]
  \centering
  \includegraphics[width=2.5in]{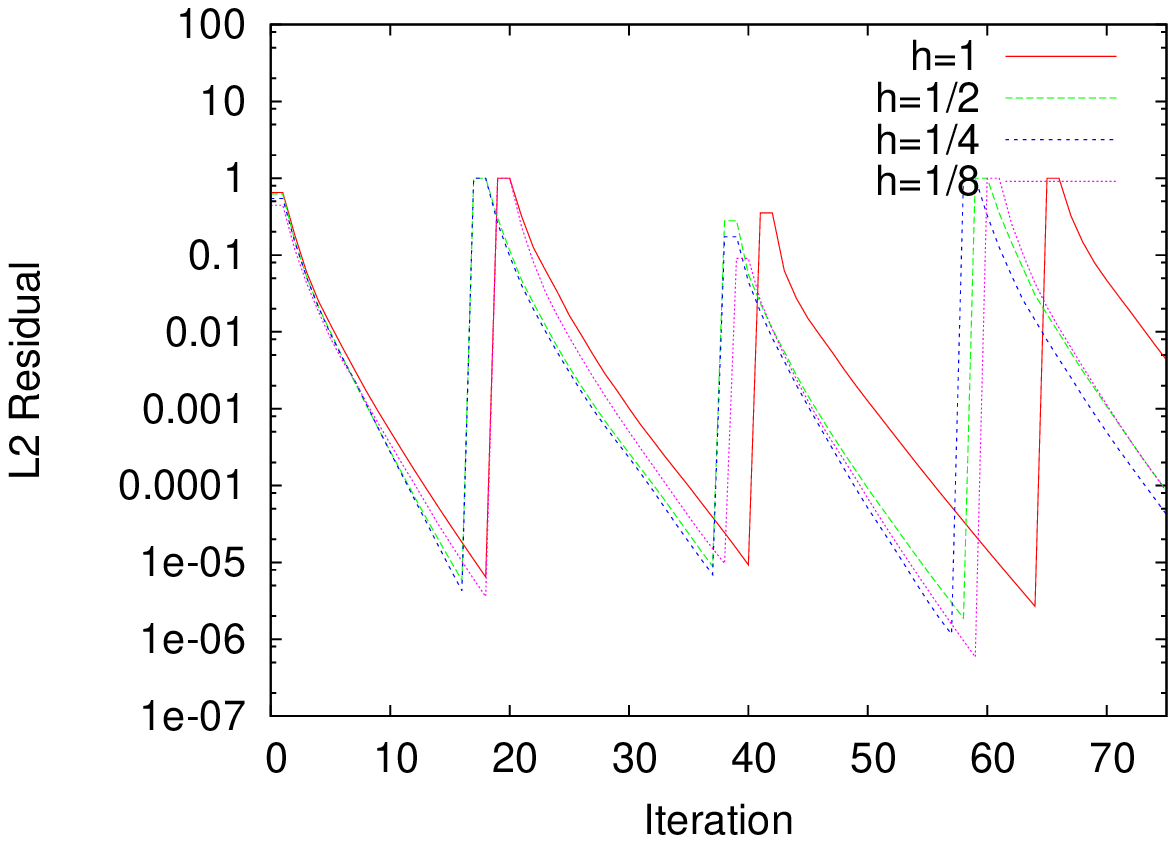}
  \includegraphics[width=2.5in]{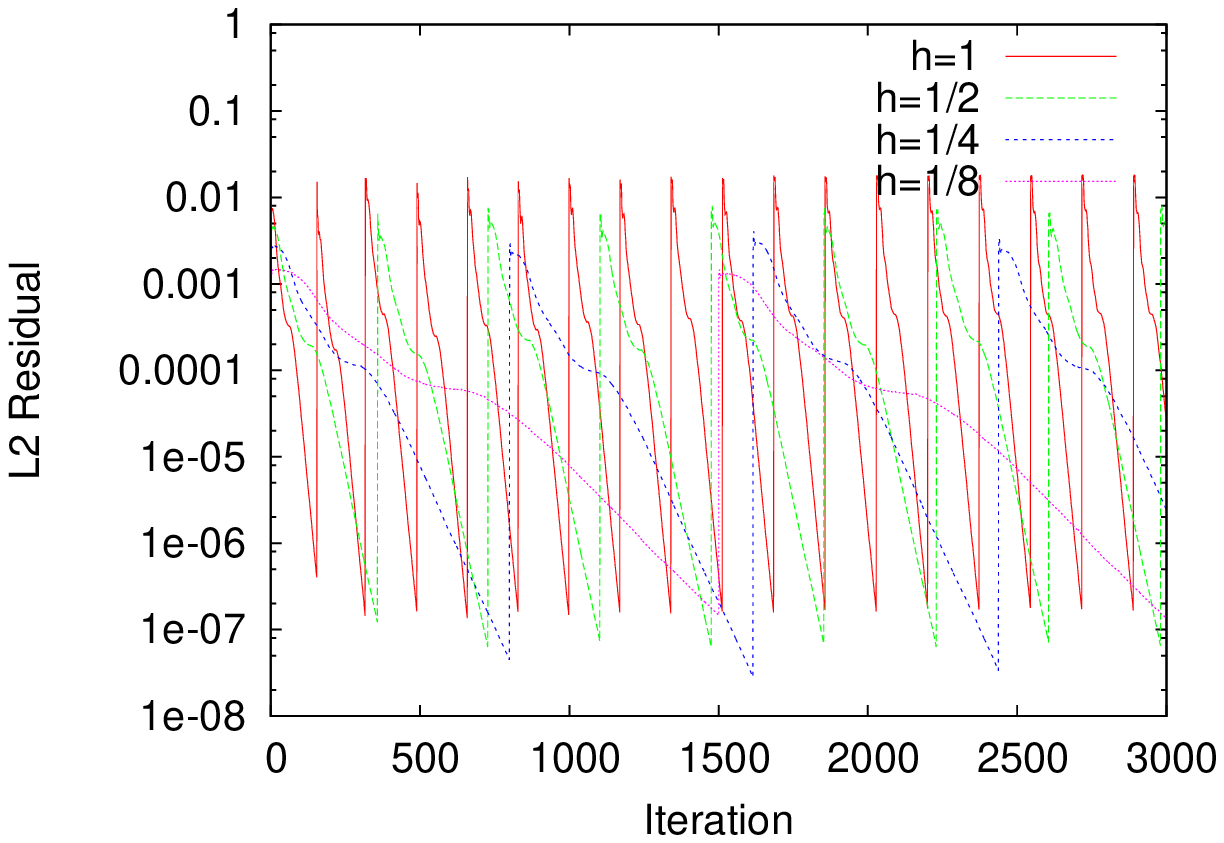}
  \caption{Kiln. Convergence of the momentum (Left) and continuity (Right) equations.}
  \label{fig:solver-cvg} 
\end{figure}

\begin{figure}[!t]
  \centering
  \includegraphics[width=0.49\columnwidth]{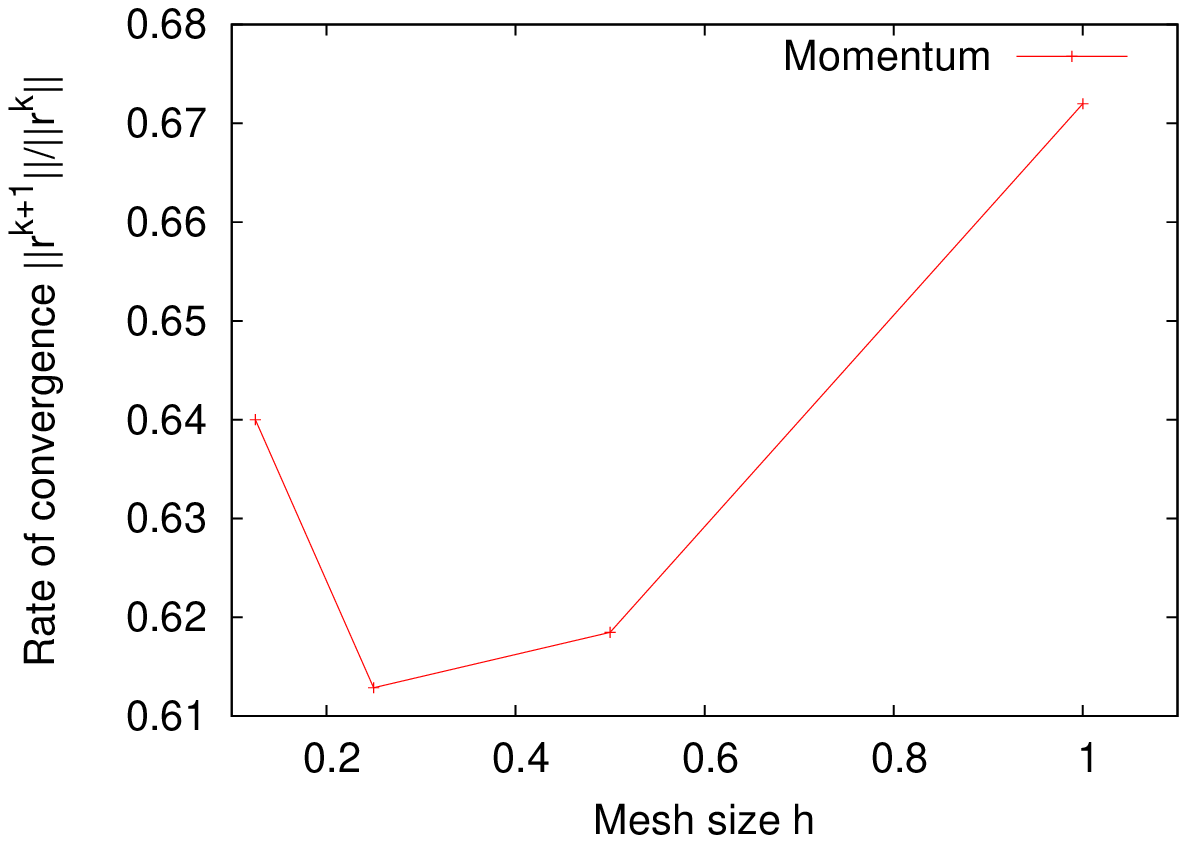} \hfill
  \includegraphics[width=0.49\columnwidth]{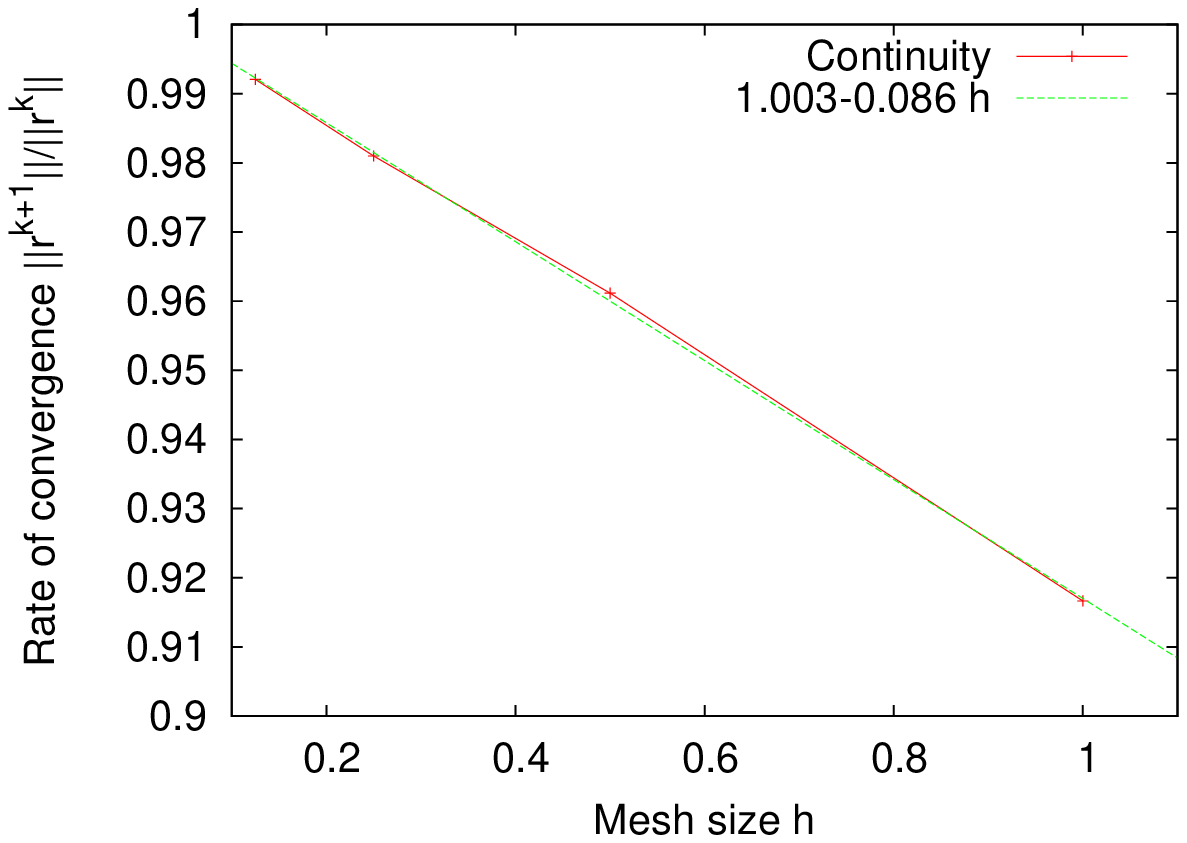}
  \caption{Kiln. rate of convergence of the momentum (Left) and continuity (Right) equations.}
  \label{fig:solver-rate} 
\end{figure}

Apart from the scalability, we present some convergence results of the solvers of the momentum (GMRES) and continuity 
equations (DCG). 
The DCG involves the solution of a coarse problem, using a direct solver, to accelerate the convergence {\em a la multigrid}, by
providing a mechanism to damp out the low frequency errors. In order to keep the number of iterations of the DCG solver
constant when refining the mesh, one can increase the number of groups, that is the size of the coarse problem. 
In the present case, the number of groups in maintained constant and is rather small for the meshes considered (=200).
Figure \ref{fig:solver-cvg} compares the converges of the first iterations for the momentum and continuity equations,
using the four meshes. 
We observe that the momentum equations converge quite rapidly and similarly for the four meshes. This is because
the time step decrease with the mesh size, increasing in this way the diagonal terms of the momentum equations, 
and leaving their condition numbers almost unchanged. This is not the case of the DCG, which converge degrades
with the mesh size. However, we observe that even with a very small number of groups, the method still
converges. In Figure \ref{fig:solver-rate} we plotted the rates of convergence of the GMRES and DCG
solvers. These figures can be useful to predict the number of iterations required to achieve a
given residual reduction according to the mesh size. In addition, we plotted an approximate linear fit to 
the rate of convergence of the DCG for this particular case.

\subsection{The electromechamical cardiac model}

The Alya Cardiac Computational Model is explained in papers such as \cite{Vazquez11Heart,lafo12}.
Simulating a heart beat is a complex, multiscale problem. This means
that many scales are coupled together covering different orders of
magnitude: from descriptions of electrical propagation, cells
arrangement into a spatial description and up to the geometry of the
cardiac chambers \cite{LeGrice01}. In \cite{Trayanova2011} the authors
review the effort to model the behavior of the heart during the last
decades, from the molecular point of view to the anatomical level of
the organ. On the one hand, electrophysiological models at the cell
level consider Ca-based activation. Mechanical models representing the
deformation of the tissue are based on protein interaction like
actin-myosin and events in cardiac myofilaments and single cells. On
the other hand, the electromechanical solution at organ level requires coupling
electrical and mechanical components and can include some of the
cellular models, depending on the specific application. 
At the organ-level, the cardiac computational model requires solution of
the electrical component as a non-linear reaction-diffusion system, i.e. an excitable media model,
the mechanical component, which produces the deformation and a
coupling scheme to link both problems together. 

The electrical propagation is modelled as a diffusion equation with local anisotropy and a non-linear
term. Anisotropy is due to the fiber-like complex structure of the cardiac muscle, fibers are defined
as a nodal field coming from either mathematical modelling or a special kind of Magnetic Resonance Imaging.
Combined with the diffusion, the non-linear term produces a sharp depolarization advancing front. In 
this example, a so-called FitzHugh-Nagumo model is used.

The second Physical problem is the muscular contraction and relaxation. 
From the mechanical point of view, the myocardium is here considered compressible. The material
is hyper-elastic, with anisotropic behaviour ruled by the fiber
structure.  In this work, we use a transversally isotropic version
based on \cite{Holzapfel09} and presented in \cite{lafo12}. 
The dynamical mechanical equations are written in a total-Lagrangian formulation.
The Cauchy stress $\boldsymbol{\sigma}=J^{-1}\boldsymbol{P}\boldsymbol{F}^T$, related to the first Piola-Kirchoff $P_{iJ}$ and
the deformation gradient $F_{iJ}=\Frac{\partial x_i}{ \partial X_J}$ and its Jacobian $J$, allows to define the material model.
Stress is developed in two parts: active and passive:
\begin{equation}
\boldsymbol{\sigma} = \boldsymbol{\sigma}_{pas} + \sigma_{act}(\lambda,[Ca^{2+}]) \boldsymbol{f} \otimes \boldsymbol{f}
\end{equation}
The passive part is governed by
a transverse isotropic exponential strain energy function $W(b)$ that
relates the Cauchy stress $\sigma$ to the right Cauchy-Green deformation $b$. The passive stress is then 
\begin{eqnarray}
J \boldsymbol{\sigma}_{pas} 
= (a~e^{b(I_1-3)}-a)\boldsymbol{b} 
+2 a_f (I_4-1)e^{b_f(I_4-1)^2}\boldsymbol{{f}} \otimes \boldsymbol{{f}} \nonumber
\\ 
+ K (J-1) \boldsymbol{I}
\end{eqnarray}
The strain invariant $I_1$ represents the non-collagenous material while
strain invariant $I_4$ represents the stiffness of the muscle fibers, and $a, b, a_f ,
b_f$ are parameters to be determined experimentally. $K$ sets the compressibility. 
Vector $\boldsymbol{{f}}$ defines the fiber direction. 

Electro-mechanical action depends on ionic concentration in the tissue, being
coupling models still under development. 
The electrical component simulates the propagation of the
transmembrane potential by solving a reaction-diffusion system plus some non-linear terms. By solving
these equations, 
ion concentrations ($Ca^{2+}$, $Na^+$, $K^+$...) in the cellular membrane can be computed. 

Electro-mechanical coupling is modelled as follows. Cardiac mechanical deformation is the result of the active tension generated
by the myocytes. The model includes passive and active properties of
the myocardium. It assumes that the active stress is produced only in
the direction of the fiber and depends on the calcium concentration of
the cardiac cell, as described in papers such as \cite{NiedererHunter06}:
\begin{equation}
\sigma_{act}=\alpha \; \frac{[Ca^{2+}]^n}{[Ca^{2+}]^n + C^n_{50}} \sigma_{max}(1+\beta(\lambda_f-1)).
\end{equation}
In this equation, $C^n_{50}$, $\sigma_{max}$ and $\lambda_f$ are model parameters. We have introduced
a parameter $0<\alpha <1$ to calibrate the amount of active stress and measure its sensitivity.

\begin{figure}[!t]
  \centering
  \includegraphics[width=3.5in]{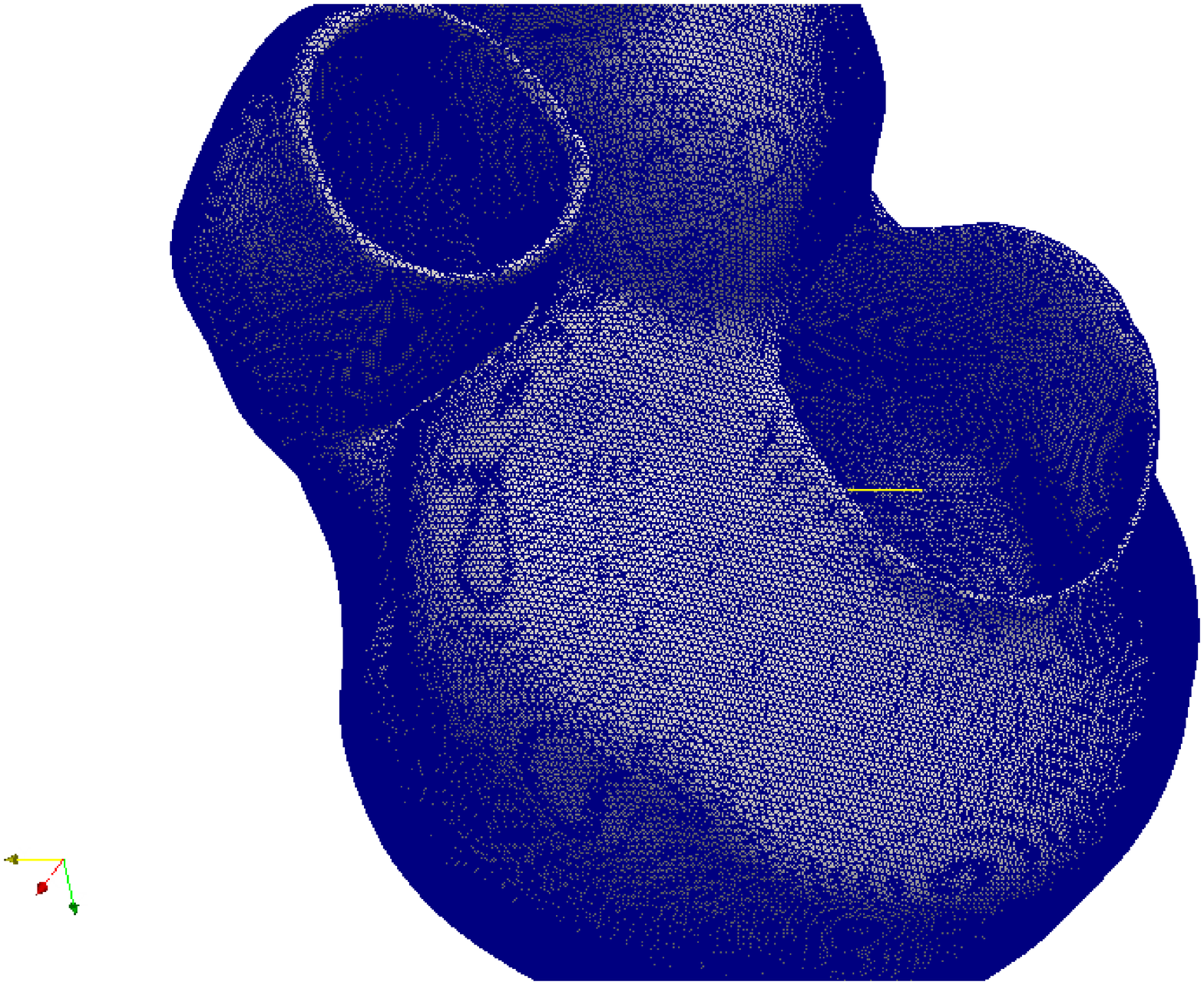}
  \includegraphics[width=3.5in]{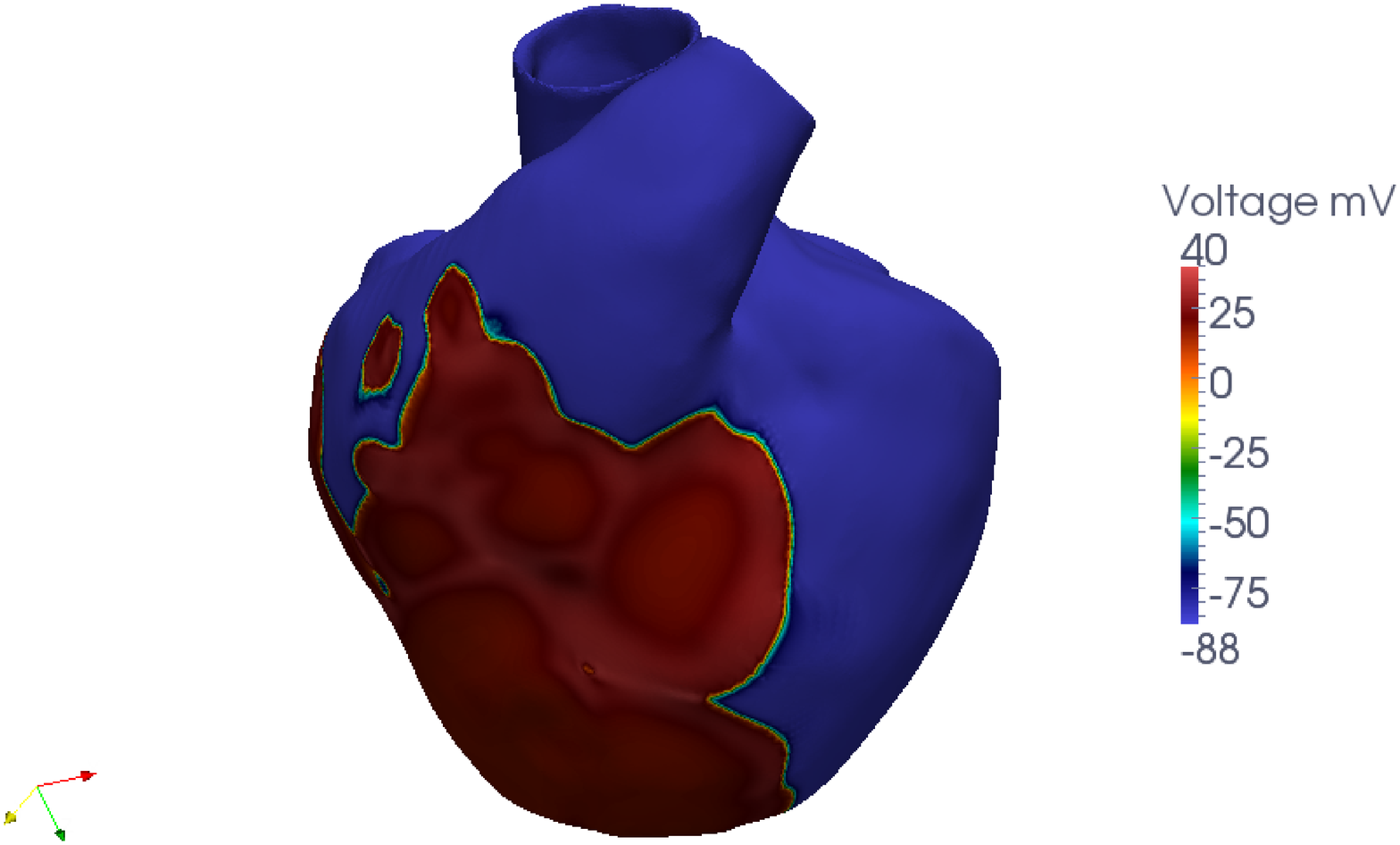}
  \caption{Heart initial mesh (top) and electrophysiology activation potential.}
  \label{fig:supercaso} 
\end{figure}

In order to capture all the required time scales, small time steps are needed. Therefore, in cardiac
mechanics simulations explicit schemes for time integration are preferred. Figure \ref{fig:supercaso} shows a snapshot
on the electromechanical propagation, closing up on the mesh. The original mesh is made of a bit more than 6M elements.
After two and three subdivision cycles following \cite{HouzeauxMultipli},  
it reaches 427 millions and 3.4 billions tetrahedra respectively. 

\begin{figure}[!t]
  \centering
  \includegraphics[width=4.5in]{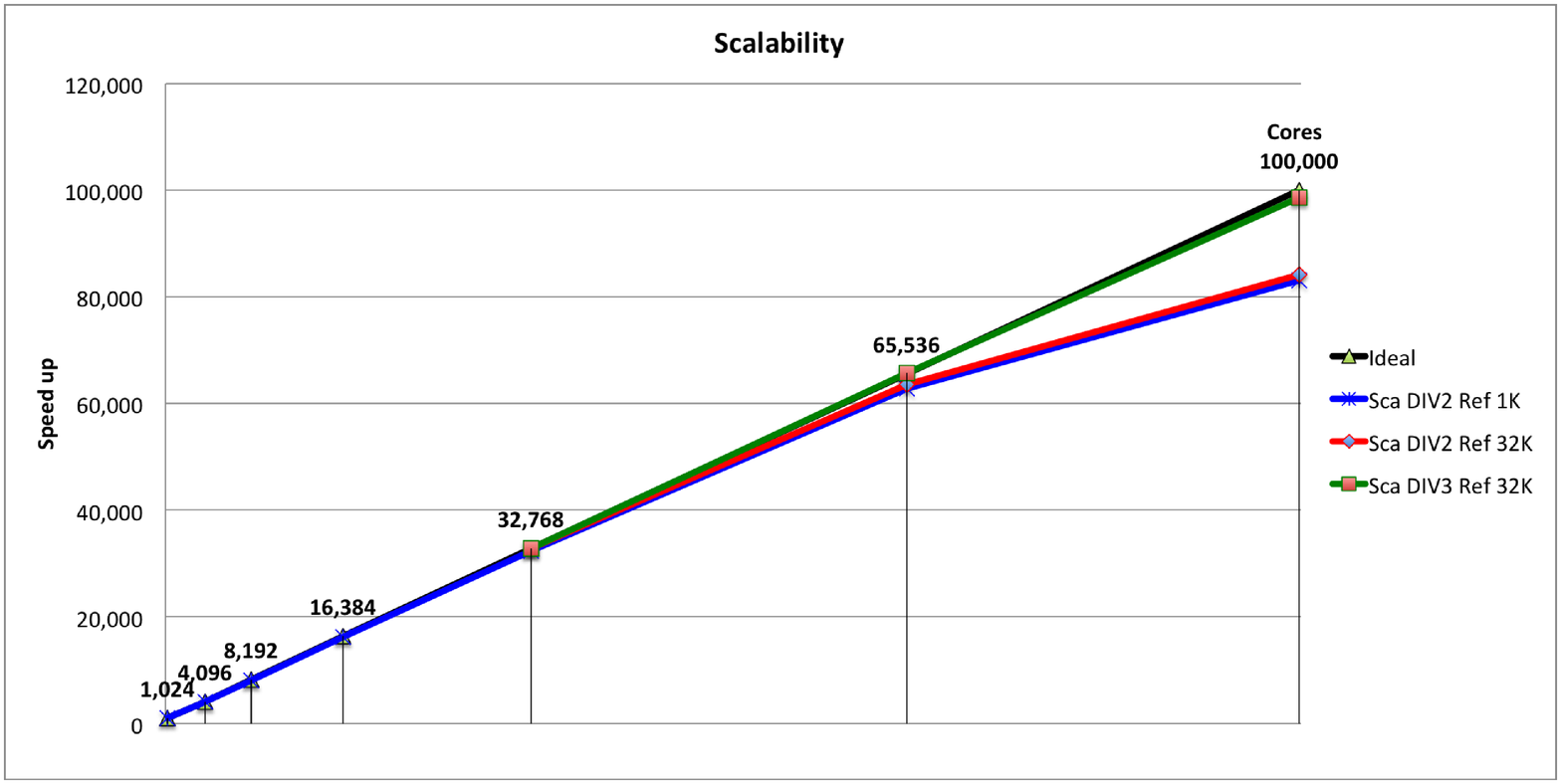}
  \vskip0.2cm
  \includegraphics[width=4.5in]{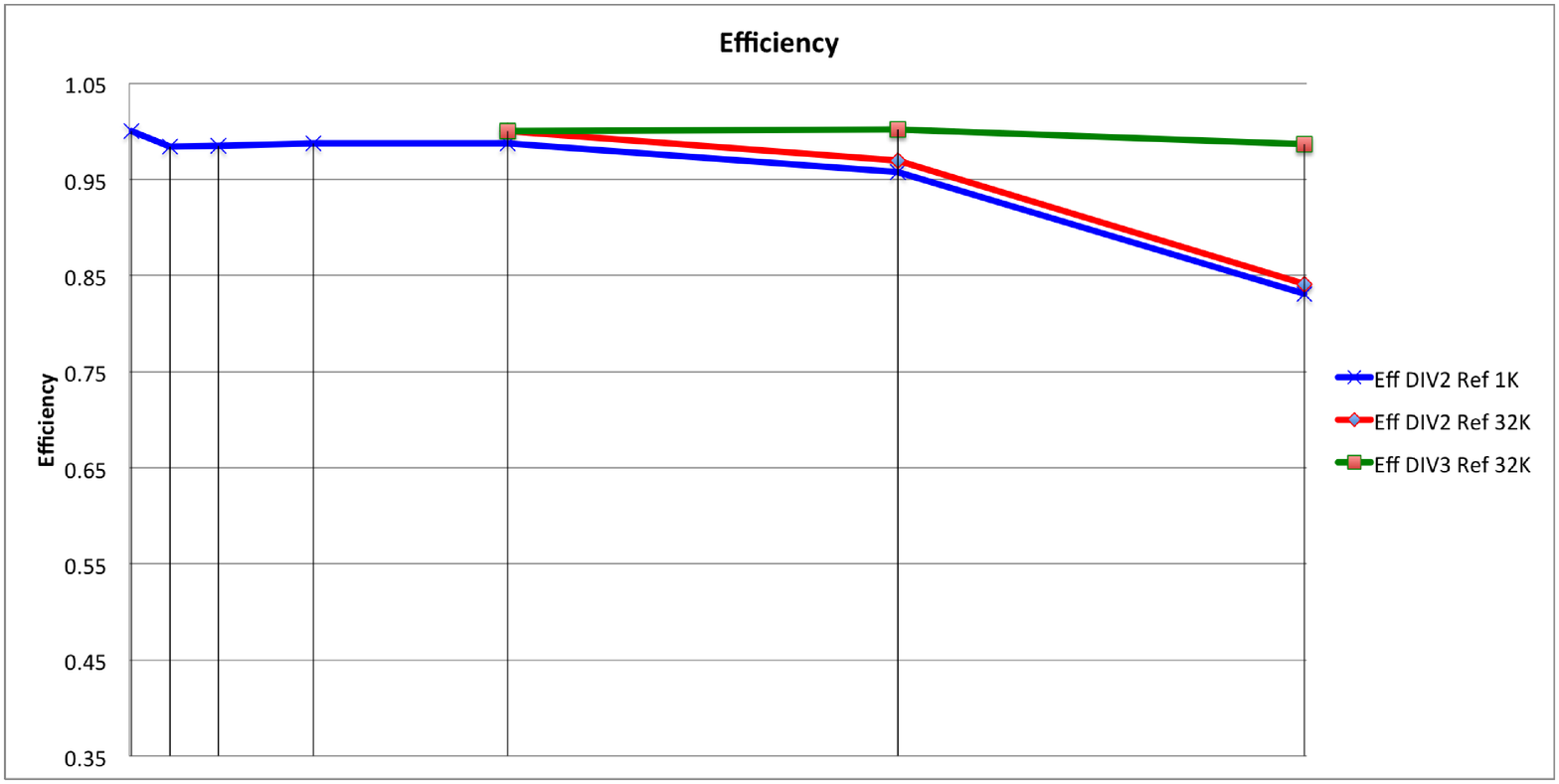}
  \vskip0.2cm
  \includegraphics[width=4.5in]{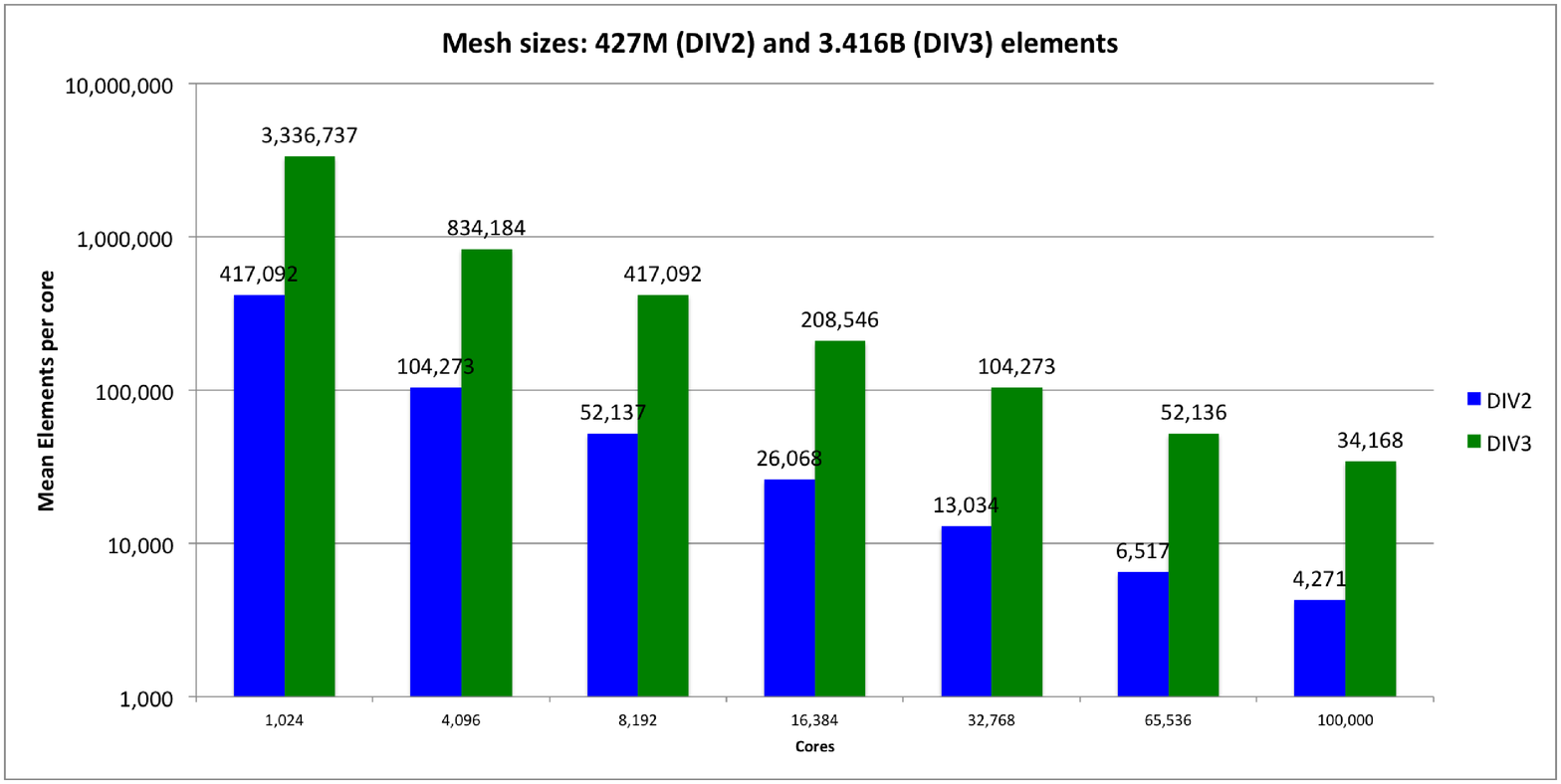}
  \caption{Scalability, efficiency and mean-elements-per-core for the cardiac electro-mechanical simulation.}
  \label{fig:supercaso-scalabi}
\end{figure}

Figure \ref{fig:kiln-scalabi} shows the strong scalability and efficiency for the cardiac electromechanical model. Again, 
the plots show the total scalability, measured summing up the CPU times for the Physical problems solved, namely
electrophysiology and solid mechanics. As in the kiln example, 
we show here the results for two meshes: 427M (called DIV2) and 3.416B elements (called DIV3). Using the same
approach as above, for DIV2 (labelled ``DIV2 Ref 1K'')
the scalability
is measured all the way from 1024 up to 100K cores. In this case, the sweet spot is lower than 4K elements per core.
The reason is that, being the scheme explicit, communication needs per time step are much lower. 
As in the kiln example, for DIV3, i.e. the largest mesh, we run the last three points of the plot, 32768, 65536 and 100000 cores,
using as the scalability normalizing value the CPU time obtained for 32768 (labelled ``DIV3 Ref 32K''). 
We have added the scalability and efficiency plots for DIV2, but now normalizing with 32768 instead of 1024 (labelled ``DIV2 Ref 32K''). 
Again and as expected, ``DIV2 Ref 32K'' is very close to a translation upwards of ``DIV2 Ref 1K''. On the other hand, ``DIV3 Ref 32K''
presents a much better scalability and efficiency, with a sustained large efficiency up to 100.000 cores.

\section{Conclusions and Future Lines}
In this paper, we have presented the simulation strategy of Alya, a multi-physics solver designed to run efficiently on 
tens of thousands of processors, especially well-suited for simulations in the engineering realm. The NCSA's 
sustained peta-scale system of Blue Waters has shown its potential with large-scale Alya runs maintaining unprecedentedly 
high parallel efficiency on 100,000 cores, clearly demonstrating the feasibility of petascale (and potentially exascale) 
computing in engineering.   

The chosen examples cover the largest possible
range of features such a code should have. We solved coupled multi-physics incompressible fluid mechanics, combustion and thermal 
flow, non-linear solid mechanics and excitable media. We simulated meshes up to billions
of elements. We used both explicit and implicit sehemes, showing scalability plots for both cases 
and analyzing solver convergence for the implicit ones. We 
used non-structured hybrid meshes combined with a mesh sub-division strategy. Finally, we concluded that 
very large scale coupled multi-physics simulations are feasible and efficient on tens of thousands of cores when 
using codes like Alya in systems like Blue Waters.

Such large-scale problems represent a completely new territory, revealing new
issues.
Solution strategies must be adapted to take advantage of supercomputers. In the case of the Navier-Stokes equations, an
algebraic split strategy has enabled us to the use relatively classical iterative solvers with very good convergence and
parallel performances for very large unstructured and hybrid meshes. But sufficient load is necessary to keep parallel
efficiencly as high as possible. The efficiency obtained on some numerical examples gives us some lower bound estimation,
depending on the physics and numerical schemes.

Scalability is the first step. Next, solver convergence must be analyzed. We show
in this study at what extent Alya strategy is correct, but there are plenty of 
issues still to be treated. One key problem is postprocessing. For problems
of this size, tools such as HDF5 are very important, but their behaviour on these
grounds must be analyzed. 

The next steps in Alya roadmap towards exascale will take different directions.
We will analyze scalability 
of very large problems, but
now when multi-physics are separated, like in the case of Fluid-Structure Interaction. 
Considering accelerators, let us remark that in recent years, 
substantial efforts were undertaken to adapt computational sparse methods for evolving 
GPU systems. We plan to test GPU-based solver and mathematical libraries with Alya such as CuBLAS 
\cite{cublas} and Paralution \cite{paralution} on XK7 nodes of Blue Waters. We also plan to test a massively parallel 
direct solver library WSMP \cite{wsmp} as a preconditioner for the iterative solvers in Alaya. WSMP has 
shown enough scalability and robustness to perform with multi-million equation problem size on many 
thousands of cores \cite{guptakoric}. Besides, Alya has been tested in Intel Xeon Phi systems with sustained
scalability specially using the MPI parallelization paradigm and virtually no effort in porting. Further research
in these two lines will be carried out and reported.


\section*{Acknowledgment}
The authors would like to thank the following fellow researchers and institutions:
\begin{itemize}
\item The Private Sector Program at NCSA and the Blue Waters sustained-petascale 
computing project-supported by the National Science Foundation (award number OCI 07-25070) and the state of Illinois.
\item Denis Doorly and Alister Bates (Imperial College London, UK), collaborators of the airways study. Part of this work
was financed by European PRACE Type B/C projects.
\item The heart geometry was provided by Dr. A. Berruezo (Hospital Clínic de Barcelona) in 
collaboration with R. Sebastian (UVEG) and O. Camara (UPF), partially financed through project TIN2011-28067 from MINECO, Spain.
\item Part of the cardiac model development was financed by the
grant SEV-2011-00067 of Severo Ochoa Program, awarded by the Spanish Government.
\item Part of the kiln model development was financed by the European Commission in the framework of the FP7 Collaborative project 
``Advanced Technologies for the Production of Cement and Clean Aggregates from Construction and Demolition Waste (C2CA)'', 
Grant Agreement No 265189.
\end{itemize}



\bibliographystyle{IEEEtran}
\bibliography{IEEEabrv,biblio}

\end{document}